\documentclass[usenatbib]{mn2e}

\usepackage{times}
\usepackage{lscape}
\usepackage{subfigure}
\usepackage[normalem]{ulem}
\usepackage{lscape}
\input{psfig.sty}

\def\gsim{\mathrel{\raise0.35ex\hbox{$\scriptstyle >$}\kern-0.6em\lower0.40ex\hbox{{$\scriptstyle \sim$}}}}
\def\lsim{\mathrel{\raise0.35ex\hbox{$\scriptstyle <$}\kern-0.6em\lower0.40ex\hbox{{$\scriptstyle \sim$}}}}

\def\apj{ApJ}
\def\apjl{ApJL}
\def\mnras{MNRAS}

\def\aaps{AAPS}
\def\araa{ARAA}
\def\aap{A\&AP}
\def\aj{AJ}
\def\apjs{APJS}

\makeatother

\title[A MUSE \& KMOS kinematic survey of galaxies from
  $z$\,=\,0.3--1.7]{Angular momentum evolution of galaxies over the
  past 10\,Gyr: A MUSE and KMOS dynamical survey of 400 star-forming
  galaxies from $z$\,=\,0.3--1.7}

\author[Swinbank et al.]
{ \parbox[h]{\textwidth}{
A.\,M.\ Swinbank,$^{\! 1,2}$
C.\ M.\, Harrison,$^{1,2}$
J.\, Trayford,$^{1,2}$
M.\, Schaller,$^{1,2}$
Ian Smail,$^{1,2}$
J.\, Schaye,$^{3}$
T.\, Theuns,$^{1,2}$
R.\, Smit,$^{1,2}$
D.\, M.\, Alexander,$^{1,2}$
R.\, Bacon,$^{4}$
R.\ G.\, Bower,$^{1,2}$
T.\, Contini,$^{5,6}$
R.\ A.\, Crain,$^{7}$
C.\ de Breuck,$^{8}$
R.\, Decarli$^{9}$,
B.\ Epinat,$^{5,6,10}$
M.\, Fumagalli,$^{1,2}$
M.\ Furlong,$^{1,2}$
A.\, Galametz,$^{11}$
H.\ L.\, Johnson,$^{1,2}$
C.\, Lagos,$^{12,13,14}$
J.\, Richard,$^{4}$
J.\ Vernet,$^{8}$,
R.\ M.\, Sharples,$^{1,2}$
D.\ Sobral,$^{15}$
\& J.\ P.\, Stott$^{1,2,16}$
}
\vspace*{6pt}\\
$^1$Institute for Computational Cosmology, Durham University, South Road, Durham DH1 3LE UK\\
$^2$Center for Extra-galactic Astronomy,  Durham University, South Road, Durham DH1 3LE UK\\
$^3$Leiden Observatory, Leiden University, PO Box 9513, NL-2300 RA Leiden, Netherlands\\
$^4$CRAL, Observatoire de Lyon, Universite Lyon 1, 9 Avenue Ch. Andre, F-69561 Saint Genis Laval Cedex, France\\
$^5$IRAP, Institut de Recherche en Astrophysique et Planetologie, CNRS, 14, avenue Edouard Belin, F-31400 Toulouse, France\\
$^6$Universite de Toulouse, UPS-OMP, Toulouse, France\\
$^7$Astrophysics Research Institute, Liverpool John Moores University, 146 Brownlow Hill, Liverpool L3 5RF, UK\\
$^8$European Southern Observatory, Karl Schwarzschild Stra{\ss}e 2, 85748, Garching, Germany \\
$^9$Max-Planck Institut f\"ur Astronomie, K\"onigstuhl 17, D-69117, Heidelberg, Germany\\
$^{10}$Aix Marseille Universit\'e, CNRS, LAM, Laboratoire d'Astrophysique de Marseille, UMR 7326, 13388 Marseille, France\\
$^{11}$Max-Planck-Institut fur Extraterrestrische Physik, D-85741 Garching, Germany\\
$^{12}$International Centre for Radio Astronomy Research (ICRAR), M468, University of Western Australia, 35 Stirling Hwy, Crawley, WA 6009, Australia\\
$^{13}$Australian Research Council Centre of Excellence for All-sky Astrophysics (CAASTRO), 44 Rosehill Street Redfern, NSW 2016, Australia\\
$^{14}$Kavli Institute for Theoretical Physics, Kohn Hall, University of California, Santa Barbara, CA 93106, United States\\
$^{15}$Department of Physics, Lancaster University, Lancaster, LA1 4BY, UK\\
$^{16}$Sub-department of Astrophysics, University of Oxford, Denys Wilkinson Building, Keble Road, Oxford OX1 3RH, UK\\
$^{*}$ email: a.m.swinbank@durham.ac.uk\\
}

\begin{document}

\pagerange{} \pubyear{2016}
\volume{}

\maketitle 

\begin{abstract}
We present a MUSE and KMOS dynamical study 405 star-forming galaxies
at redshift $z$\,=\,0.28--1.65 (median redshift $\bar z$\,=\,0.84).
Our sample are representative of star-forming, main-sequence galaxies,
with star-formation rates of SFR\,=\,0.1--30\,M$_{\odot}$\,yr$^{-1}$
and stellar masses M$_{\star}$\,=\,10$^8$--10$^{11}$\,M$_{\odot}$.
For 49\,$\pm$\,4\% of our sample, the dynamics suggest rotational
support, 24\,$\pm$\,3\% are unresolved systems and 5\,$\pm$\,2\%
appear to be early-stage major mergers with components on 8--30\,kpc
scales.  The remaining 22\,$\pm$\,5\% appear to be dynamically
complex, irregular (or face-on systems).  For galaxies whose dynamics
suggest rotational support, we derive inclination corrected rotational
velocities and show these systems lie on a similar scaling between
stellar mass and specific angular momentum as local spirals with
$j_\star$\,=\,$J$\,/\,$M_\star\propto M_\star^{2/3}$ but with a
redshift evolution that scales as
$j_\star\,\propto$\,M$_\star^{2/3}(1+z)^{-1}$.  We also identify a
correlation between specific angular momentum and disk stability such
that galaxies with the highest specific angular momentum
(log($j_\star$\,/\,M$_\star^{2/3}$)\,$>$\,2.5) are the most stable,
with Toomre $Q$\,=\,1.10\,$\pm$\,0.18, compared to
$Q$\,=\,0.53\,$\pm$\,0.22 for galaxies with
log($j_\star$\,/\,M$_\star^{2/3}$)\,$<$\,2.5.  At a fixed mass, the
\emph{HST} morphologies of galaxies with the highest specific angular
momentum resemble spiral galaxies, whilst those with low specific
angular momentum are morphologically complex and dominated by several
bright star-forming regions.  This suggests that angular momentum
plays a major role in defining the stability of gas disks: at
$z\sim$\,1, massive galaxies that have disks with low specific angular
momentum, are globally unstable, clumpy and turbulent systems.  In
contrast, galaxies with high specific angular have evolved in to
stable disks with spiral structure where star formation is a local
(rather than global) process.
\end{abstract}

\begin{keywords}
  galaxies: evolution --- galaxies: high-redshift --- galaxies: dynamics
\end{keywords}

\section{Introduction}

%
%
\begin{figure*}
  \centerline{\psfig{file=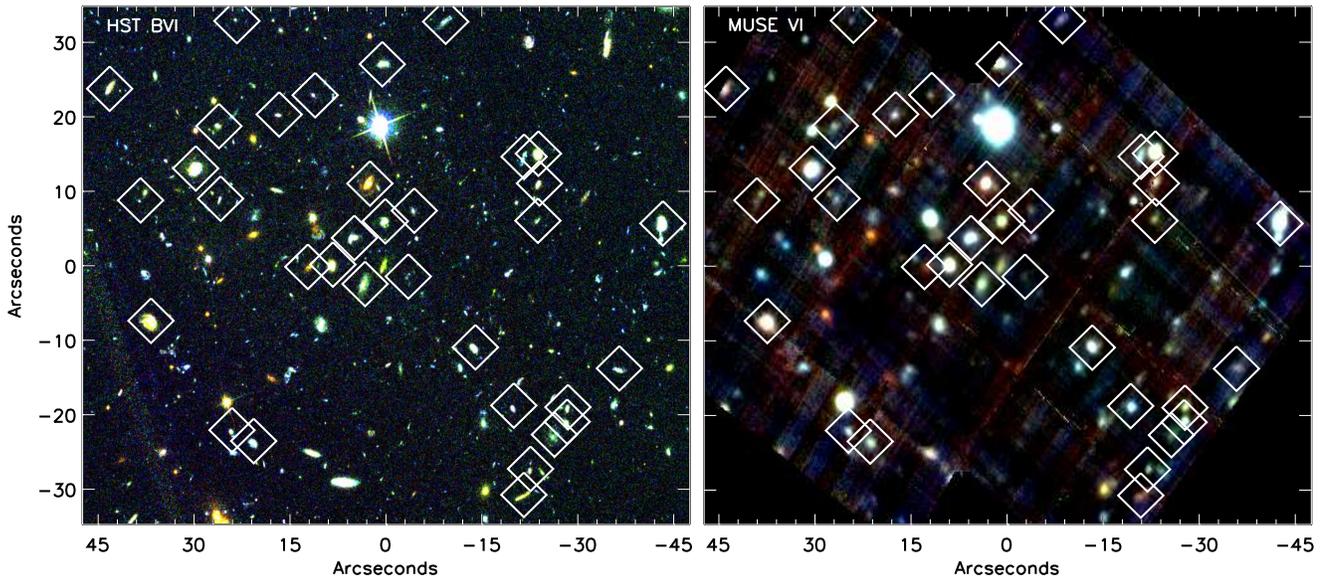,width=7in,angle=0}}
\caption{$HST$ and MUSE images for one of our survey fields,
  TN\,J1338$-$13 which contains a $z$\,=\,4.4 radio galaxy --
  underlining the fact that our survey of the foreground galaxy
  population is unbiased. {\it Left:} $HST$ $BVI$-band colour image.
  The [O{\sc ii}] emitters identified from this field are also marked
  by open symbols.  {\it Center:} MUSE $VI$-band colour image of the
  cube generated from three equal wavelength ranges.  The [O{\sc ii}]
  emitters are again marked.  Each image is centered with (0,0) at
  $\alpha$: 13\,38\,26.1, $\delta$: $-$19\,42\,30.5 with North up and
  East left.}
\label{fig:colimg}
\end{figure*}

Identifying the dominant physical processes that were responsible for
the formation of the Hubble sequence has been one of the major goals
of galaxy formation for decades
\citep[][]{Roberts63,Gallagher84,Sandage86}.  Morphological surveys of
high-redshift galaxies, in particular utilizing the high angular
resolution of the \emph{Hubble Space Telescope; (HST)} have suggested
that only at $z\sim$\,1.5 did the Hubble sequence begin to emerge
\citep[e.g.\ ][]{Bell04,Conselice11}, with the spirals and ellipticals
becoming as common as peculiar galaxies
\citep[e.g.\ ][]{Buitrago13,Mortlock13}.
However, galaxy morphologies reflect the complex (non-linear)
processes of gas accretion, baryonic dissipation, star formation and
morphological transformation that have occured during the history of
the galaxy.  Furthermore, morphological studies of high-redshift
galaxies are subject to K-corrections and structured dust obscuration,
which complicates their interpretation.

The more fundamental physical properties of galaxies are their mass,
energy and angular momentum, since these are related to the amount of
material in a galaxy, the linear size and the rotational velocity.  As
originally suggested by \citet{Sandage70}, the Hubble sequence of
galaxy morphologies appears to follow a sequence of increasing angular
momentum at a fixed mass \cite[e.g.\ ][]{Fall83,Fall13,Obreschkow14}.
One route to identifying the processes responsible for the formation
of disks is therefore to measure the evolution of the mass, size and
dynamics (and hence angular momentum) of galaxy disks with cosmic time
-- properties which are more closely related to the underlying dark
matter halo.

In the cold dark matter paradigm, baryonic disks form at the centers
of dark matter halos.  As dark matter halos grow early in their
formation history, they acquire angular momentum ($J$) as a result of
large scale tidal torques.  The angular momentum acquired has strong
mass dependence, with $J\propto M_{\rm halo}^{5/3}$
\citep[e.g.\ ][]{Catelan96}.  Although the halos acquire angular
momentum, the centrifugal support of the baryons and dark matter
within the virial radius is small.  Indeed, whether calculated through
linear theory or via $N$-body simulations, the ``spin'' (which defines
the ratio of the halo angular speed to that required for the halo to
be entirely centrifugally supported) follows approximately a
log-normal distribution with average value $\lambda_{\rm
  DM}$\,=\,0.035 \citep{Bett07}.  This quantity is invariant to
cosmological parameters, time, mass or environment
\citep[e.g.\ ][]{Barnes87,Steinmetz95,Cole96}.

As the gas collapses within the halo, the baryons can both lose and
gain angular momentum between the virial radius and disk scale.  If
the baryons are dynamically cold, they fall inwards, weakly conserving
specific angular momentum.  Although the spin of the baryon at the
virial radius is small, by the time they reach $\sim$\,2--10\,kpc (the
``size'' of a disk), they form a centrifugally supported disk which
follows an exponential mass profile \citep[e.g.\ ][]{Fall83,Mo98}.
Here, ``weakly conserved'' is within a factor of two, and indeed,
observational studies suggest that late-type spiral disks have a spin
of $\lambda_{\rm disk}'$\,=\,0.025; \citep[e.g.\ ][]{Courteau97},
suggesting that that that only $\sim$\,30\% of the initial baryonic
angular momentum is lost due to viscous angular momentum
redistribution and selective gas losses which occurs as the galaxy
disks forms \citep[e.g.\ ][]{Burkert09}.

In contrast, if the baryons do not make it in to the disk, are
redistributed (e.g.\ due to mergers), or blown out of the galaxy due
to winds, then the spin of the disk is much lower than that of the
halo.  Indeed, the fraction of the initial halo angular momentum that
is lost must be as high as $\sim$\,90\% for early-type and elliptical
galaxies (at the same stellar mass as spirals; \citealt{Bertola75}),
with Sa and S0 galaxies in between the extremes of late-type spiral-
and elliptical- galaxies \citep[e.g.\ ][]{Romanowsky12}.

Numerical models have suggested that most of the angular momentum
transfer occurs at epochs ealier than $z\sim1$, after which the
baryonic disks gain sufficient angular momentum to stabilise
themselves \citep{Dekel09b,Ceverino10,Obreschkow15,Lagos16}.  For
example, \citet{Danovich15} use identify four dominant phases of
angular momentum exchange that dominate this process: linear tidal
torques on the gas beyond and through the virial radius; angular
momentum transport through the halo; and dissipation and disk
instabilities, outflows in the disk itself.  These processes can
increase and decrease the specific angular momentum of the disk as it
forms, although they eventually ``conspire'' to produce disks that
have a similar spin distribution as the parent dark matter halo.

Measuring the processes that control the internal redistribution of
angular momentum in high-redshift disks is observationally demanding.
However, on galaxy scales (i.e.\ $\sim$\,2--10\,kpc), observations
suggest redshift evolution according to
$j_\star$\,=\,J$_\star$\,/\,M$_\star\propto$(1+$z$)$^n$ with
$n\sim$\,$-$1.5, at least out to $z\sim$\,2
\citep[e.g.\ ][]{Obreschkow15,Burkert16}.  Recently, \citet{Burkert16}
exploited the {\sc kmos$^{\rm 3D}$} survey of $z\sim$\,1--2.5
star-forming galaxies at to infer the angular momentum distribution of
baryonic disks, finding that their spin is is broadly consistent the
dark matter halos, with $\lambda\sim$\,0.037 with a dispersion
($\sigma_{\rm log\lambda}\sim$\,0.2).  The lack of correlation between
the ``spin'' ($j_{\rm disk}$\,/\,$j_{\rm DM}$) and the stellar
densities of high-redshift galaxies also suggests that the
redistribution of the angular momentum within the disks is the
dominant process that leads to compactation (i.e. bulge formation;
Burkert et al.\ 2016; Tadaki et al.\ 2016).  Taken together,
these results suggest that angular momentum in high redshift disks
plays a dominant role in ``crystalising'' the Hubble sequence of
galaxy morphologies.

In this paper, we investigate how the angular momentum and spin of
baryonic disks evolves with redshift by measuring the dynamics of a
large, representative sample of star-forming galaxies between
$z\sim$\,0.28--1.65 as observed with the KMOS and MUSE integral field
spectrographs.  We aim to measure the angular momentum of the stars
and gas in large and representative samples of high-redshift galaxies.
Only now, with the capabilities of sensitive, multi-deployable (or
wide-area) integral field spectrographs, such as MUSE and KMOS, is
this becoming possible
\citep[e.g.\ ][]{Bacon15,Wisnioski15,Stott16,Burkert16}.  We use our
data to investigate how the mass, size, rotational velocity of galaxy
disks evolves with cosmic time.  As well as providing constraints on
the processes which shape the Hubble sequence, the evolution of the
angular momentum and stellar mass provides a novel approach to test
galaxy formation models since these values reflect the initial
conditions of their host halos, merging, and the prescriptions that
describe the processes of gas accretion, star formation and feedback,
all of which can strongly effect the angular momentum of the baryonic
disk.

In \S\ref{sec:obs} we describe the observations and data reduction.
In \S\ref{sec:analysis} we describe the analysis used to derive
stellar masses, galaxy sizes, inclinations, and dynamical properties.
In \S\ref{sec:discussion} we combine the stellar masses, sizes and
dynamics to measure the redshift evolution of the angular momentum of
galaxies.  We also compare our results to hydro-dynamical simulations.
In \S\ref{sec:conclusions} we give our conclusions.  Throughout the
paper, we use a cosmology with $\Omega_{\Lambda}$\,=\,0.73,
$\Omega_{m}$\,=\,0.27, and H$_{0}$\,=\,72\,km\,s$^{-1}$\,Mpc$^{-1}$.
In this cosmology a spatial resolution of 0.7$''$ corresponds to a
physical scale of 5.2\,kpc at $z$\,=\,0.84 (the median redshift of our
survey).  All quoted magnitudes are on the AB system and we adopt a
Chabrier IMF throughout.

\section{Observations and Data Reduction}
\label{sec:obs}

The observations for this program were acquired from a series of
programs (commissioning, guaranteed time and open-time projects; see
Table~1) with the new Multi-Unit Spectroscopic Explorer (MUSE;
\citealt{Bacon10SPIE,Bacon15}) and $K$-band Multi-Object Spectrograph
(KMOS; \citealt{Sharples04}) on the ESO Very Large Telescope (VLT).
Here, we describe the observations and data reduction, and discuss how
the properties (star-formation rates and stellar masses) of the
galaxies in our sample compare to the ``main-sequence'' population.

\subsection{MUSE Observations}
\label{sec:muse_obs}

As part of the commissioning and science verification of the MUSE
spectrograph, observations of fifteen ``extra-galactic'' fields were
taken between 2014 February and 2015 February.  The science targets of
these programs include ``blank'' field studies (e.g.\ observations of
the {Hubble Ultra-Deep Field}; \citealt{Bacon15}), as well as
high-redshift ($z>$\,2) galaxies, quasars and galaxy clusters
(e.g.\ Fig.~\ref{fig:colimg}) \citep[see
  also][]{Husband15,Richard15,Contini16}.  The wavelength coverage of
MUSE (4770--9300\,\AA\, in its standard configuration) allows us to
serendipitously identify [O{\sc ii}] emitters between
$z\sim$\,0.3--1.5 in these fields and so to study the dynamics of
star-forming galaxies over this redshift range.  We exploit these
observation to construct a sample of star-forming galaxies, selected
via their [O{\sc ii}] emission.  The program IDs, pointing centers,
exposure times, seeing FWHM (as measured from stars in the continuum
images) for all of the MUSE pointings are given in Table~1.  We also
supplement these data with [O{\sc ii}] emitters from MUSE observations
from two open-time projects (both of whose primary science goals are
also to detect and resolve the properties of $z>$\,3 galaxies/QSOs;
Table~1).  The median exposure time for each of these fields is
12\,ks, but ranges from 5.4--107.5\,ks.  In total, the MUSE survey
area exploited here is $\sim$\,20\,arcmin$^2$ with a total integration
time of 89\,hours.

The MUSE IFU provides full spectral coverage spanning
4770--9300\,\AA\, and a contiguous field of view of
60\,$''\times$\,60\,$''$, with a spatial sampling of 0.2$''$\,/\,pixel
and a spectral resolution of
$R$\,=\,$\lambda$\,/\,$\Delta\lambda$\,=\,3500 at
$\lambda$\,=\,7000\AA\, (the wavelength of the [O{\sc ii}] at the
median redshift of our sample) -- sufficient to resolve the [O{\sc
    ii}]$\lambda\lambda$3726.2,3728.9 emission line doublet.  In all
cases, each 1\,hour observing block was split in to a number of
sub-exposures (typically 600, 1200, or 1800\,seconds) with small
(2$''$) dithers between exposures to account for bad pixels.  All
observations were carried out in dark time, good sky transparency.
The average $V$-band seeing for the observations was 0.7$''$
(Table~1).

To reduce the data, we use the MUSE {\sc esorex} pipeline which
extracts, wavelength calibrates, flat-fields the spectra
and forms each datacube.  In all of the data taken after August 2014,
each 1\,hr science observation was interspersed with a flat-field to
improve the slice-by-slice flat field (illumination) effects.  Sky
subtraction was performed on each sub-exposure by identifying and
subtracting the sky emission using blank areas of sky at each
wavelength slice, and the final mosaics were then constructed using an
average with a 3-$\sigma$ clip to reject cosmic rays, using point
sources in each (wavelength collapsed) image to register the cubes.
Flux calibration was carried out using observations of known standard
stars at similar airmass and were taken immediately before or after
the science observations.  In each case we confirmed the flux
calibration by measuring the flux density of stars with known
photometry in the MUSE science field.

%
%
\begin{table*}
\begin{center}
\caption{Observing logs}
\begin{tabular}{lcccrrr}
\hline
\hline
Field Name      &  PID            &  RA              &   Dec                & $t_{\rm exp}$        &  seeing        & 3-$\sigma$ SB limit  \\
                &                 & (J2000)          & (J2000)              & ($ks$)             &   ($''$)       &                      \\
\hline
{\bf MUSE:}\\
  J0210-0555    &  060.A-9302     &  02:10:39.43    &   $-$05:56:41.28     &  9.9               &    1.08        &  9.1              \\
  J0224-0002    &  094.A-0141     &  02:24:35.10    &   $-$00:02:16.00     &  14.4              &    0.70        &  11.0             \\
  J0958+1202    &  094.A-0280     &  09:58:52.34    &   +12:02:45.00       &  11.2              &    0.80        &  15.2             \\
  COSMOS-M1     &  060.A-9100     &  10:00:44.26    &   +02:07:56.91       &  17.0              &    0.90        &  5.3              \\
  COSMOS-M2     &  060.A-9100     &  10:01:10.57    &   +02:04:10.60       &  12.6              &    1.0         &  6.3              \\
  TNJ1338       &  060.A-9318     &  13:38:25.28    &   $-$19:42:34.56     &  32.0              &    0.75        &  4.1              \\
  J1616+0459    &  060.A-9323     &  16:16:36.96    &   +04:59:34.30       &  7.0               &    0.90        &  7.6              \\
  J2031-4037    &  060.A-9100     &  20:31:54.52    &   $-$40:37:21.62     &  37.7              &    0.83        &  5.2              \\
  J2033-4723    &  060.A-9306     &  20:33:42.23    &   $-$47:23:43.69     &  7.9               &    0.85        &  7.4              \\
  J2102-3535    &  060.A-9331     &  21:02:44.97    &   $-$35:53:09.31     &  11.9              &    1.00        &  6.2              \\
  J2132-3353    &  060.A-9334     &  21:32:38.97    &   $-$33:53:01.72     &  6.5               &    0.70        &  13.6             \\
  J2139-0824    &  060.A-9325     &  21:39:11.86    &   $-$38:24:26.14     &  7.4               &    0.80        &  5.7              \\
  J2217+1417    &  060.A-9326     &  22:17:20.89    &   +14:17:57.01       &  8.1               &    0.80        &  4.9              \\
  J2217+0012    &  095.A-0570     &  22:17:25.01    &   +00:12:36.50       &  12.0              &    0.69        &  6.0              \\
  HDFS-M2       &  060.A-9338     &  22:32:52.71    &   $-$60:32:07.30     &  11.2              &    0.90        &  7.3              \\
  HDFS-M1       &  060.A-9100     &  22:32:55.54    &   $-$60:33:48.64     &  107.5             &    0.80        &  2.8              \\
  J2329-0301    &  060.A-9321     &  23:29:08.27    &   $-$03:01:58.80     &  5.7               &    0.80        &  5.6              \\
\hline                                                                                                              
    {\bf KMOS:}\\
  COSMOS-K1      &  095.A-0748    &  09:59:33.54    & +02:18:00.43        &  16.2               &   0.70         &  22.5               \\
  SSA22          &  060.A-9460    &  22:19:30.45    & +00:38:53.34        &  7.2                &   0.72         &  31.2               \\
  SSA22          &  060.A-9460    &  22:19:41.15    & +00:23:16.65        &  7.2                &   0.70         &  33.7               \\
\hline
\label{table:obs}
\end{tabular}
\end{center}
\noindent{\footnotesize Notes: RA and Dec denote the field centers.
  The seeing is measured from stars in the field of view (MUSE) or
  from a star placed on one of the IFUs (KMOS). The units of the
  surface brightness limit are
  $\times$10$^{-19}$\,erg\,s$^{-1}$\,cm$^{-2}$\,arcsec$^{-2}$.  The
  reduced MUSE datacubes for these fields available at:
  http://astro.dur.ac.uk/$\sim$ams/MUSEcubes/ \\}
\end{table*}

To identify [O{\sc ii}] emitters in the cubes, we construct and coadd
$V$- and $I$-band continuum images from each cube by collapsing the
cubes over the wavelength ranges $\lambda$\,=\,4770--7050\AA\, and
$\lambda$\,=\,7050--9300\AA\ respectively.  We then use {\sc
  sextractor} \citep{Bertin96} to identify all of the $>$4\,$\sigma$
continuum sources in the ``detection'' images.  For each continuum
source, we extract a 5\,$\times$\,5$''$ sub-cube (centered on each
continuum source) and search both the one and two-dimensional spectra
for emission lines.  At this resolution, the [O{\sc ii}] doublet is
resolved and so trivially differentiated from other emission lines,
such as Ly$\alpha$, [O{\sc iii}]\,4959,5007 or H$\alpha$+[N{\sc
    ii}]\,6548,6583.  In cases where an emission line is identified,
we measure the wavelength, $x$\,/\,$y$ (pixel) position and RA\,/\,Dec
of the galaxy.  Since we are interested in resolved dynamics, we only
include galaxies where the [O{\sc ii}] emission line is detected above
5\,$\sigma$ in the one dimensional spectrum.  To ensure we do not miss
any [O{\sc ii}] emitters that do not have continuum counterparts, we
also remove all of the continuum sources from each cube by masking a
5$''$ diameter region centered on the continuum counterpart, and
search the remaining cube for [O{\sc ii}] emitters.  We do not
find any additional [O{\sc ii}]-emitting galaxies where the integrated
[O{\sc ii}] flux is detected above a signal-to-noise of 5 (i.e.\ all
of the bright [O{\sc ii}] emitters in our sample have at least a
4\,$\sigma$ detection in continuum).

In Fig.~\ref{fig:colimg} we show a \emph{HST} $BVI$-band colour image
of one of our target fields, TNJ\,1338, along with a colour image
generated from the 32\,ks MUSE exposure.  The blue, green and red
channels are generated from equal width wavelength ranges between
4770--9300\,AA\, in the MUSE cube.  In both panels we identify all of
the [O{\sc ii}] emitters.  In this single field alone, there are 33
resolved [O{\sc ii}] emitters.  

\begin{figure*}
  \centerline{\psfig{file=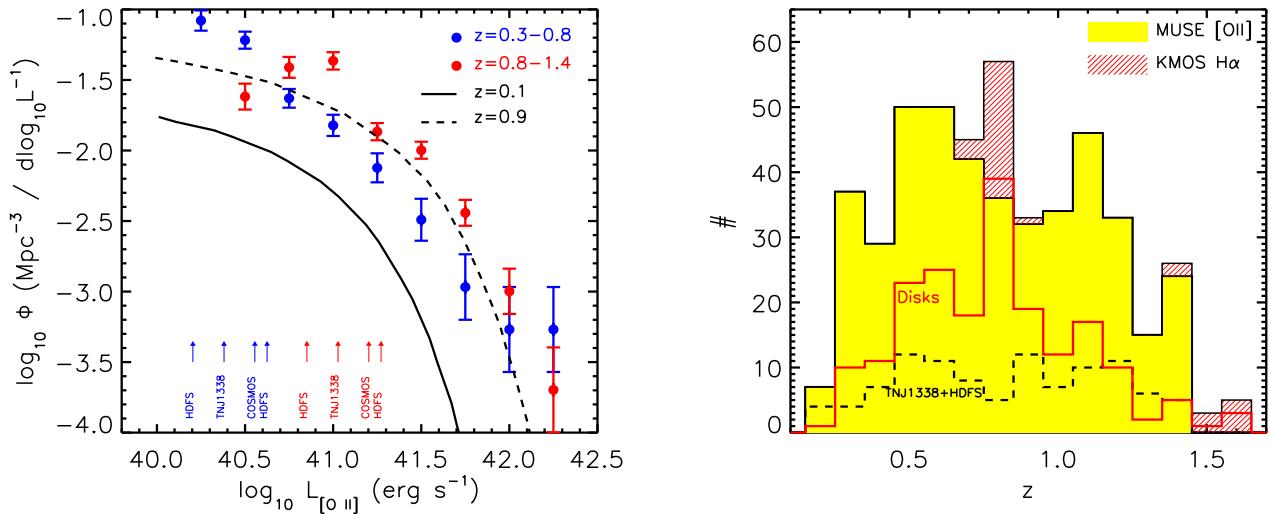,width=7in,angle=90}}
  \caption{{\it Left:} [O{\sc ii}] luminosity function for the
    star-forming galaxies in our sample from the 18 MUSE IFU
    pointings.  We split the sample in two redshift bins,
    $z$\,=\,0.3--0.8 and $z$\,=\,0.8--1.4.  The arrows on the plot
    denote luminosity limits for four of the fields in the MUSE sample
    (which span the complete range of depths).  To baseline these
    results, we overlay the [O{\sc ii}] luminosity function at
    $z$\,=\,0 from SDSS \citep{Ciadrdullo13} which shows that there is
    strong evolution in L$^{\star}_{\rm [OII]}$ from $z\sim$\,0 to
    $z\sim$\,0.5.  This evolution is also seen in other [O{\sc ii}]
    surveys \citep[e.g.\ ][]{Ly07,Khostovan15}.  {\it Right:} The
    redshift distribution of the [O{\sc ii}] and H$\alpha$ emitters in
    our MUSE and KMOS samples.  Our sample has a median redshift of
    $z$\,=\,0.84 and a full redshift range of $z$\,=\,0.28--1.67.
    Since the MUSE observations have a wide range of exposure times,
    from 5.7--107.5\,ks, we overlay the redshift distribution of the
    [O{\sc ii}] emitters in the two deepest fields, HDFS and TNJ\,1338,
    to highlight that the highest-redshift galaxies are not dominated
    by the deepest observations.  We also overlay the redshift
    distribution of the galaxies classified as ``rotationally
    supported'' (i.e.\ disks).}
  \label{fig:Nz}
\end{figure*}

From all 17 MUSE fields considered in this analysis, we identify a
total of 431 [O{\sc ii}] emitters with emission line fluxes ranging
from 0.1--170\,$\times$\,10$^{-17}$\,erg\,s$^{-1}$\,cm$^{-2}$ with a
median flux of 3\,$\times$\,10$^{-17}$\,erg\,s$^{-1}$\,cm$^{-2}$ and a
median redshift of $z$\,=\,0.84 (Fig.~\ref{fig:Nz}).

Before discussing the resolved properties of these galaxies, we first
test how our [O{\sc ii}]-selected sample compares to other [O{\sc ii}]
surveys at similar redshifts.  We calculate the [O{\sc ii}] luminosity
of each galaxy and in Fig.~\ref{fig:Nz} show the [O{\sc ii}]
luminosity function in two redshift bins ($z$\,=\,0.3--0.8 and
$z$\,=\,0.8--1.4).  In both redshift bins, we account for the
incompleteness caused by the exposure time differences between fields.
We highlight the luminosity limits for four of the fields which span
the whole range of depths in our survey.  This figure shows that the
[O{\sc ii}] luminosity function evolves strongly with redshift, with
$L^{\star}$ evolving from log$_{\rm
  10}$($L^{\star}$[erg\,s$^{-1}$\,cm$^{-2}$])\,=\,41.06\,$\pm$\,0.17
at $z$\,=\,0 to log$_{\rm
  10}$($L^{\star}$[erg\,s$^{-1}$\,cm$^{-2}$])\,=\,41.5\,$\pm$\,0.20
and log$_{\rm
  10}$($L^{\star}$[erg\,s$^{-1}$\,cm$^{-2}$])\,=\,41.7\,$\pm$\,0.22 at
$z$\,=\,1.4 \citep[see also][]{Ly07,Khostovan15}.  The same evolution
has also been seen in at UV wavelengths \citep{Oesch10} and in
H$\alpha$ emission \citep[e.g.\ ][]{Sobral13}.

%
%

\subsection{KMOS Observations}
\label{sec:kmos_obs}

We also include observations of the redshifted H$\alpha$ in 46
$z\sim$0.8--1.7 galaxies from three well-studied extra-galactic
fields.  Two of these fields are taken from an H$\alpha$-selected
sample at $z$\,=\,0.84 from the KMOS--Hi-$z$ emission line survey
(KMOS-HiZELS; \citealt{Geach08,Sobral09,Sobral13}) and are discussed
in \citet{Sobral13b,Sobral15} and \citet{Stott14}.  Briefly,
observations of 29 H$\alpha$-selected galaxies were taken between 2013
June and 2013 July using KMOS with the $YJ$-band filter as part of the
KMOS science verification programme.  The near--infrared KMOS IFU
comprises 24 IFUs, each of size 2.8\,$\times$\,2.8$''$ sampled at
0.2$''$ which can be deployed across a 7-arcmin diameter patrol field.
The total exposure time was 7.2\,ks per pixel, and we used
object-sky-object observing sequences, with one IFU from each of the
three KMOS spectrographs placed on sky to monitor OH variations.

Further KMOS observations were also obtained between 2015 April 25 and
April 27 as the first part of a 20-night KMOS guaranteed time
programme aimed at resolving the dynamics of 300 mass-selected
galaxies at $z\sim$\,1.2--1.7.  Seventeen galaxies were selected from
photometric catalogs of the COSMOS field.  We initially selected
targets in the redshift range $z$\,=\,1.3--1.7 and brighter than
$K_{\rm AB}$\,=\,22 (a limit designed to ensure we obtain sufficient
signal-to-noise per resolution element to spatially resolve the
galaxies; see \citealt{Stott16} for details).  To ensure that the
H$\alpha$ emission is bright enough to detect {\it and} spatially
resolve with KMOS, we pre-screened the targets using the
\emph{Magellan Multi-object Infra-Red Spectrograph} {\sc (mmirs)} to
search for and measure the H$\alpha$ flux of each target, and then
carried out follow-up observations with KMOS of those galaxies with
H$\alpha$ fluxes brighter than
5\,$\times$\,10$^{-17}$\,erg\,s$^{-1}$\,cm$^{-2}$.  These KMOS
observations were carried out using the $H$-band filter, which has a
spectral resolution of $R$\,=\,$\lambda$\,/\,$\Delta\lambda$\,=\,4000.
We used object-sky-object sequences, with one of the IFUs placed on a
star to monitor the PSF and one IFU on blank sky to measure OH
variations.  The total exposure time was 16.2\,ks (split in to three
5.4\,ks OBs, with 600\,s sub-exposures).  Data reduction was performed
using the {\sc spark} pipeline with additional sky-subtraction and
mosaicing carried out using customized routines.  We note that a
similar dymamical\,/\,angular momemtum analysis of the $\sim$\,800
galaxies at $z\sim$\,1 from the KROSS survey are presented in
\citet{Harrison17}.

\subsection{Final Sample}
\label{sec:finalsample}
Combining the two KMOS samples, in total there are 41\,/\,46
H$\alpha$-emitting galaxies suitable for this analysis (i.e. H$\alpha$
detected above a S\,/\,N\,$>$\,5 in the collapsed, one-dimensional
spectrum).  From our MUSE sample of 431 galaxies, 67 of the faintest
[O{\sc ii}] emitters are only detected above a S\,/\,N\,=\,5 when
integrating a 1\,$\times$\,1$''$ region, and so no longer considered
in the following analysis, leaving us with a sample of 364 [O{\sc ii}]
emitters for which we can measure resolved dynamics.  Together, the
MUSE and KMOS sample used in the following analysis comprises 405
galaxies with a redshift range $z$\,=\,0.28--1.63.  We show the
redshift distribution for the full sample in Fig.~\ref{fig:Nz}.

\section{Analysis}
\label{sec:analysis}

With the sample of 405 emission-line galaxies in our survey fields,
the first step is to characterize the integrated properties of the
galaxies.  In the following, we investigate the spectral energy
distributions, stellar masses and star formation rates, sizes,
dynamics, and their connection with the galaxy morphology, and we put
our findings in the context of our knowledge of the general galaxy
population at these redshifts.  We first discuss their stellar masses.

\subsection{Spectral Energy Distributions and Stellar Masses}
\label{sec:SEDs}

The majority of the MUSE and KMOS fields in our sample have excellent
supporting optical\,/\,near- and mid-infrared imaging, and so to infer
the stellar masses and star formation rates for the galaxies in our
sample, we construct the spectral energy distributions for each
galaxy.  In most cases, we exploit archival \emph{HST}, Subaru,
\emph{Spitzer}\,/\,IRAC, UKIRT\,/\,WFCAM and\,/\,or VLT\,/\,Hawk-I
imaging.  In the optical\,/\,near-infrared imaging, we measure 2$''$
aperture photometry, whilst in the IRAC 3.6\,/\,4.5-$\mu$m bands we
use 5$''$ apertures (and apply appropriate aperture corrections based
on the PSF in each case).  We list all of the properties for each
galaxy, and show their broad-band SEDs in Table~A1.  We use {\sc
  hyper}-{\it z} \citep{Bolzonella00} to fit the photometry of each
galaxy at the known redshift, allowing a range of star formation
histories from late to early types and redennings of A$_{\rm
  V}$\,=\,0--3 in steps of $\Delta$A$_{\rm V}$\,=\,0.2 and a Calzetti
dust reddening curve \citep{Calzetti00}.  In cases of non detections,
we adopt a 3$\sigma$ upper limit.

We show the observed photometry and overlay the best-fit {\sc
  hyper}-{\it z} SED for all of the galaxies in our sample in Fig.~A1--A3.
Using the best-fit parameters, we then estimate the stellar mass of
each galaxy by integrating the best-fit star-formation history,
accounting for mass loss according to the {\sc starburst99} mass loss
rates \citep{Leitherer99}.  We note that we only calculate stellar
masses for galaxies that have detections in $>$3 wavebands, although
include the best SEDs for all sources in Fig.~A1--A3.  Using the stellar
masses and rest-frame $H$-band magnitudes, we derive a median
mass-to-light ratio for the full sample of M$_\star$\,/\,L$_{\rm
  H}$\,=\,0.20\,$\pm$\,0.01.  The best-fit reddening values and the
stellar masses for each galaxy are also given in Table~A1.

As a consistency check that our derived stellar masses are consistent
with those derived from other SED fitting codes, we compare our
results with \citet{Muzzin13} who derive the stellar masses of
galaxies in the COSMOS field using the {\sc easy} photometric redshift
code \citep{Brammer08EASY} with stellar mass estimated using {\sc
  fast} \cite{Kriek09}.  For the 54 [O{\sc ii}] emitting galaxies in
the COSMOS field in our sample, the stellar masses we derive are a
factor 1.19\,$\pm$\,0.06\,$\times$ higher than those derived using
{\sc fast}.  Most of this difference can be attributed to degeneracies
in the redshifts and best-fit star-formation histories.  Indeed, if we
limit the comparison to galaxies where the photometric and
spectroscopic redshifts agree within $\Delta z<$\,0.2, and where the
luminosity weighted ages also agrees to within a factor of 1.5, then
then the ratio of the stellar masses from {\sc hyper-z}\,/\,{\sc easy}
are 1.02\,$\pm$\,0.04\,$\times$.

\begin{figure}
  \centerline{\psfig{file=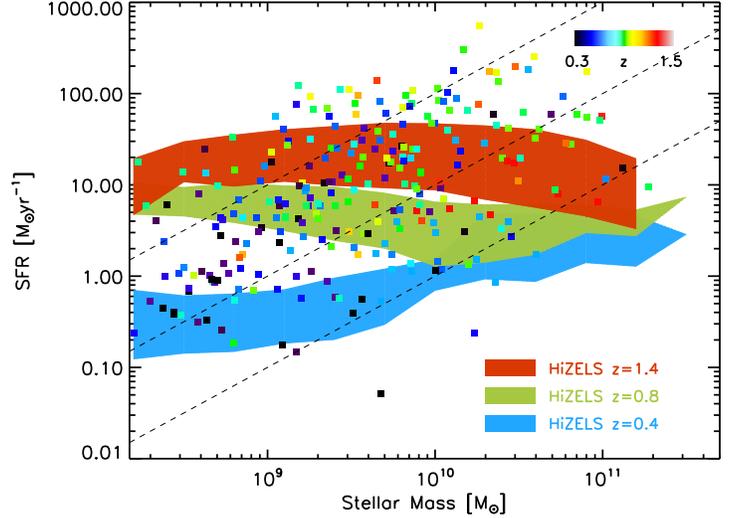,width=3.7in,angle=90}}
  \caption{Star-formation rate versus mass for the galaxies in our
    sample (with points colour-coded by redshift).  As a guide, we
    also overlay tracks of constant specific star formation rate
    (sSFR) with with sSFR\,=\,0.1, 1 and 10 Gyr$^{-1}$.  We also
    overlay the star formation rate--stellar mass relation at three
    redshift slices ($z$\,=\,0.40, 0.84 and 1.47) from the H$\alpha$
    narrow-band selected sample from HiZELS \citep{Sobral13}.  This
    shows that although the galaxies in our MUSE and KMOS samples span
    a wide range of stellar mass and star-formation rate, they are
    comparable to the general field population, with specific star
    formation rates sSFR\,$\sim$\,0.1--10\,Gyr$^{-1}$.}
  \label{fig:SSFR}
\end{figure}

To place the galaxies we have identified in the MUSE and KMOS data in
context of the general population at their respective redshifts, next
we calculate their star formation rates (and specific star formation
rates).  We first calculate the [O{\sc ii}] or H$\alpha$ emission
luminosity (L$_{\rm [OII]}$ and L$_{\rm H\alpha}$ respectively).  To
account for dust obscuration, we adopt the best-fit stellar redenning
(A$_{\rm V}$) from the stellar SED returned by from {\sc hyper-z} and
convert this to the attenuation at the wavelength of interest (A$_{\rm
  [OII]}$ or A$_{\rm H\alpha}$) using a Calzetti reddennign law;
\citealt{Calzetti00}).  Next, we assume that the the gas and stellar
phases are related by A$_{\rm
  gas}$\,=\,A$_{\star}$\,(1.9\,$-$\,0.15\,A$_{\star}$);
\citep{Wuyts13}, and then calculate the total star-formation rates
using SFR\,=\,$C$\,$\times$\,10$^{-42}$\,L$_{\sc [OII]}$\,10$^{0.4
  A_{\rm gas}}$ with $C$\,=\,0.82 and $C$\,=\,4.6 for the [O{\sc ii}]
and H$\alpha$ emitters respectively.  The star formation rates of the
galaxies in our sample range from 0.1--300\,M$_{\odot}$\,yr$^{-1}$.
In Fig.~\ref{fig:SSFR} we plot the specific star-formation rate
(sSFR\,=\,SFR\,/\,M$_{\star}$) versus stellar mass for the galaxies in
our sample.  This also shows that our sample display a wide range of
stellar masses and star-formation rates, with median and quartile
ranges of log$_{\rm
  10}$(M$_{\star}$\,/\,M$_{\odot}$)\,=\,9.4\,$\pm$\,0.9 and
SFR\,=\,4.7$_{-2.5}^{+2.2}$\,M$_{\odot}$\,yr$^{-1}$.  As a guide, in
this plot we also overlay a track of constant star formation rate with
SFR\,=\,1\,M$_{\odot}$\,yr$^{-1}$.  To compare our galaxies to the
high-redshift star-forming population, we also overlay the specific
star formation rate for $\sim$\,2500 galaxies from the HiZELS survey
which selects H$\alpha$ emitting galaxies in three narrow redshifts
slices at $z$\,=\,0.40, 0.84 and 1.47 \citep{Sobral13}.  For this
comparison, we calculate the star formation rates for the HiZELS
galaxies in an identical manner to that for our MUSE and KMOS sample.
This figure shows that the median specific star formation rate of the
galaxies in our MUSE and KMOS samples appear to be consistent with the
so-called ``main-sequence'' of star-forming galaxies at their
appropriate redshifts.

%
%

\subsection{Galaxy Sizes and Size Evolution}
\label{sec:GalaxySizes}

Next, we turn to the sizes for the galaxies in our sample.  Studies of
galaxy morphology and size, particularly from observations made with
\emph{HST}, have shown that the physical sizes of galaxies increase
with cosmic time \citep[e.g.\ ][]{Giavalisco96,Ferguson04,Oesch10}.
Indeed, late-type galaxies have continuum (stellar) half light radii
that are on average a factor $\sim$\,1.5\,$\times$ smaller at
$z\sim$\,1 than at the present day \citep{vanderwel14,Morishita14}.
As one of the primary aims of this study is to investigate the angular
momentum of the galaxy disks, the continuum sizes are an important
quantity.

\begin{figure*}
  \centerline{\psfig{file=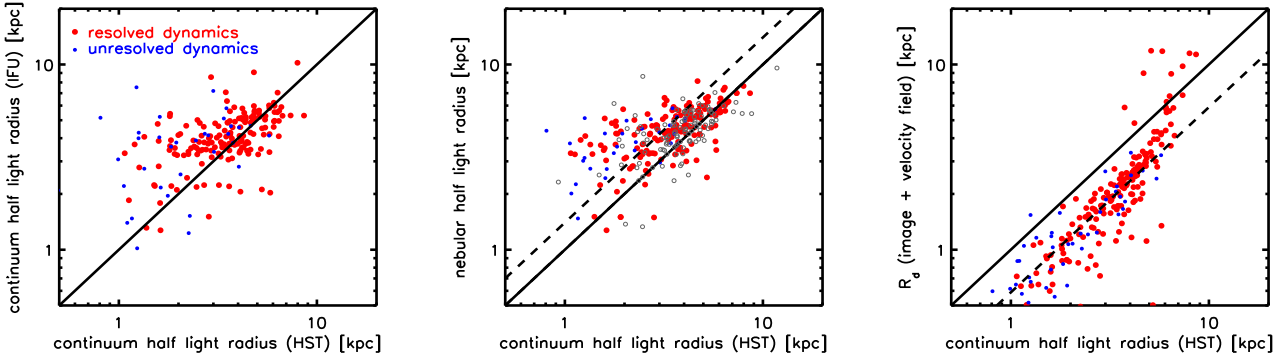,width=7in,angle=90}}
\caption{Comparison of the physical half-light radii of the galaxies
  in our sample as measured from \emph{HST} and MUSE\,/\,KMOS imaging.
  {\it Left:} Continuum half-light radii as measured from \emph{HST}
  broad-band imaging compared to those measured from the MUSE
  continuum image.  Large red points denote sources that are resolved
  by MUSE or KMOS.  Small blue point denote galaxies that are
  unresolved (or compact) in the MUSE or KMOS data.  The median ratio
  of the half-light radii is $r_{\rm HST}$\,/\,$r_{\rm
    MUSE}$\,=\,0.97\,$\pm$\,0.03 (including unresolved sources and
  deconvolved for seeing).  {\it Center:} Continuum half-light radius
  from \emph{HST} versus nebular emission half-light radius (MUSE and
  KMOS) for the galaxies in our sample from MUSE and KMOS.  The
  continuum and nebular emission line half-light radii are well
  correlated, although the nebular emission lines half-light radii are
  systematically larger than the continuum sizes, with $r_{\rm
    [OII]}$\,/\,$r_{\rm HST}$\,=\,1.18\,$\pm$\,0.03 (see also
  \citealt{Nelson15}).  Although not included in the fit, we also
  include on the plot the contuinuum size measurements from MUSE and
  KMOS as small grey points.  These increase the scatter (as expected
  from the data in the left-hand panel), although the median ratio of
  nebular emission to continuum size is unaffected if these points are
  included.  {\it Right:} Comparison of the disk scale length
  (measured from the dynamical modeling) versus the continuum
  half-light radius from \emph{HST}.  The median ratio of the
  half-light radius is larger than the disk radius by a factor $r_{\rm
    HST}$\,/\,$R_{\rm d}$\,=\,1.70\,$\pm$\,0.05, which is the
  consistent with that expected for an exponential disk.}
\label{fig:compare_sizes}
\end{figure*}

We calculate the half light-radii in both continuum and emission lines
for all galaxies in our sample.  Approximately 60\% of the galaxies in
our sample have been observed with \emph{HST} (using ACS\,/\,$BVI$
and\,/\,or WFC3\,/\,$JH$-band imaging).  Since we are interested in
the extent of the stellar light, we measure the half light radius for
each galaxy in the longest wavelength image available (usually ACS
$I-$ or WFC $H$-band).  To measure the half-light radius of each
galaxy, we first fit a two-dimensional Sersic profile to the galaxy
image to define an $x$\,/\,$y$ center and ellipticity for the galaxy,
and then measure the total flux within 1.5\,$\times$ Petrosian radius
and use the curve of growth (growing ellipses from zero to
1.5\,$\times$ Petrosian aperture) to measure the half-light radius.  A
significant fraction of our sample do not have observations with
\emph{HST} and so we also construct continuum images from the IFU
datacubes and measure the continuum size in the same way (deconvolving
for the PSF).  In Fig.~\ref{fig:compare_sizes} we compare the
half-light radius of the galaxies in our sample from \emph{HST}
observations with that measured from the MUSE and KMOS continuum
images.  From this, we derive a median ratio of $r_{1/2,\rm
  HST}$\,/\,$r_{1/2,\rm MUSE}$\,=\,0.97\,$\pm$\,0.03 with a scatter of
30\% (including unresolved sources in both cases).  

For each galaxy in our sample, we also construct a
continuum-subtracted narrow-band [O{\sc ii}] or H$\alpha$ emission
line image (using 200\AA\, on either size of the emission line to
define the continuum) and use the same technique to measure the
half-light radius of the nebular emission.  The continuum and nebular
emission line half light radii (and their errors) for each galaxy are
given in Table~A1.  As Fig.~\ref{fig:compare_sizes} shows, the nebular
emission is more extended that the continuum with $r_{1/2,\rm
  [OII]}$\,/\,$r_{1/2,\rm HST}$\,=\,1.18\,$\pm$\,0.03.  This is
consistent with recent results from the 3-D \emph{HST} survey
demonstrates that the nebular emission from $\sim$\,$L^{\star}$
galaxies at $z\sim$\,1 tends to be systematically more extended than
the stellar continuum (with weak dependence on mass;
\citealt{Nelson15}).

We also compare the continuum half light radius with the disk scale
length, $R_{\rm d}$ (see \S~\ref{sec:DynModel}).  From the data, we
measure a $r_{\rm 1/2,HST}$\,/\,$R_{\rm d}$\,=\,1.70\,$\pm$\,0.05.
For a galaxy with an exponential light profile, the half light radii
and disk scale length are related by $r_{\rm 1/2}$\,=\,1.68\,$R_{\rm
  d}$, which is consistent with our measurements (and we overlay this
relation in Fig.~\ref{fig:compare_sizes}).  In Fig.~\ref{fig:rh_z} we
plot the evolution of the half-light radii (in kpc) of the nebular
emission with redshift for the galaxies in our sample which shows that
the nebular emission half-light radii are consistent with similar
recent measurements of galaxy sizes from \emph{HST} \citep{Nelson15},
and a factor $\sim$\,1.5\,$\times$ smaller than late-type galaxies at
$z$\,=\,0.

From the full sample of [O{\sc ii}] or H$\alpha$ emitters, the spatial
extent of the nebular emission of 75\% of the sample are spatially
resolved beyond the seeing, with little\,/\,no dependence on redshift,
although the unresolved sources unsurprisingly tend to have lower
stellar masses (median M$_{\rm \star}^{\rm
  unresolved}$\,=\,1.0\,$\pm$\,0.5\,$\times$\,10$^9$\,M$_{\odot}$
compared to median M$_{\rm \star}^{\rm
  resolved}$\,=\,3\,$\pm$\,1\,$\times$\,10$^9$\,M$_{\odot}$).

%
%

\subsection{Resolved Dynamics}
\label{sec:ResolvedDynamics}

Next, we derive the velocity fields and line-of-sight velocity
dispersion maps for the galaxies in our sample.  The two-dimensional
dynamics are critical for our analysis since the circular velocity,
which we will use to determine the angular momentum in
\S~\ref{sec:discussion}, must be taken from the rotation curve at a
scale radius.  The observed circular velocity of the galaxy also
depends on the disk inclination, which can be determined using either
the imaging, or dynamics, or both.

\begin{figure*}
  \centerline{\psfig{file=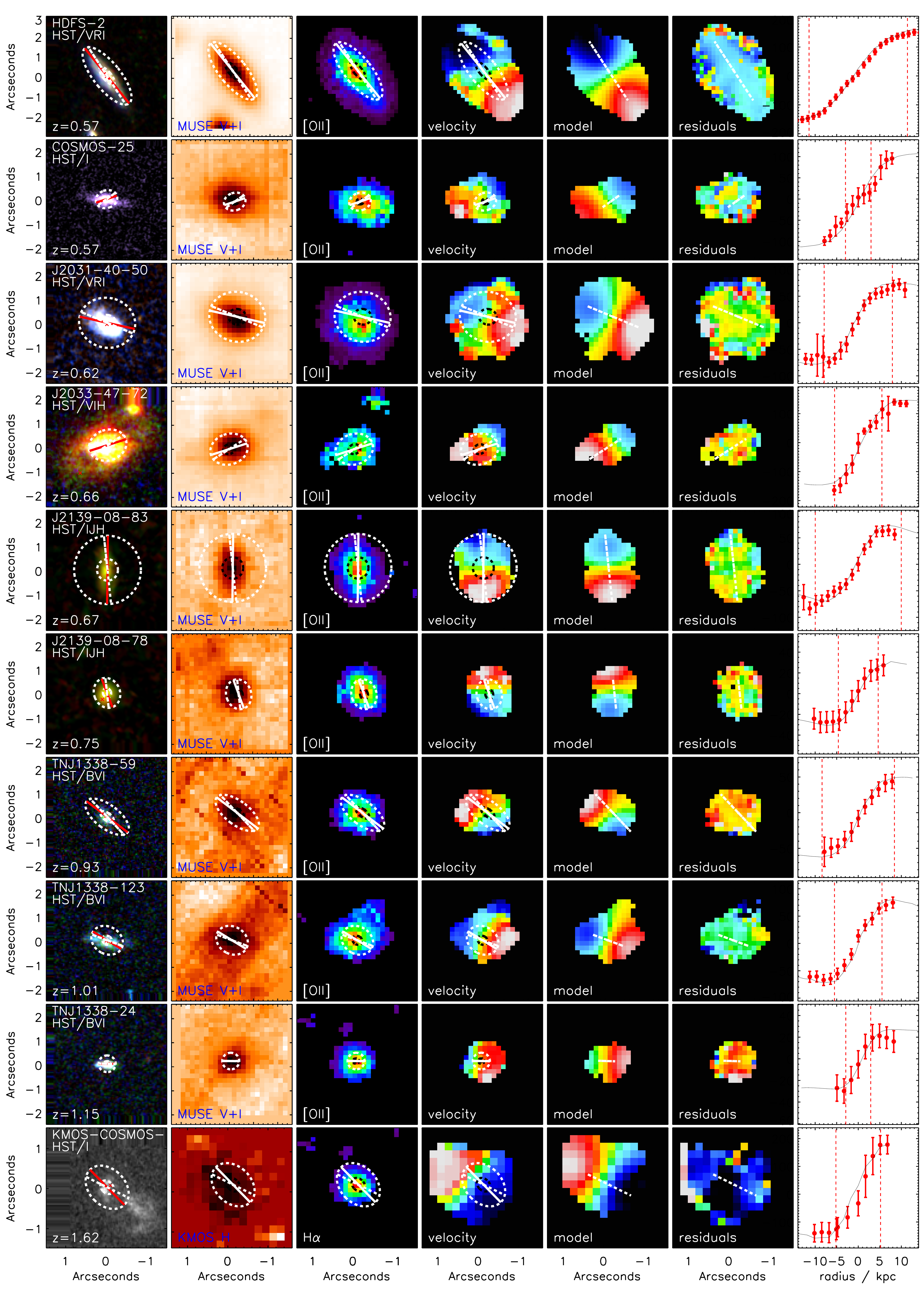,width=6in,angle=0}}
\caption{Example images and dynamics of nine galaxies in our
  sample. {\it (a):} \emph{HST} colour image of each galaxy.  
  given in each sub-image.  The galaxies are ranked by increasing
  redshift.  The ellipses denote the disk radius (inner ellipse
  R$_{\rm d}$; outer ellipse 3\,R$_{\rm d}$).  The cross denotes the
  dynamical center of the galaxy and the white-dashed and solid red
  line show the major morphological and kinematic axes
  respectively. {\it (b):} The continuum image from the IFU
  observations (dark scale denotes high intensity).  The dashed lines
  are the same as in the first panel.  {\it (c):} Nebular emission
  line velocity field.  Dashed ellipses again show the disk radius at
  R$_{\rm d}$ and 3\,R$_{\rm d}$ (the colour scale is set by the
  range shown in the final panel).  {\it (d):} Best-fit
  two-dimensional dynamical model for each galaxy.  In this panel, the
  cross and dashed line denote the dynamical center and major
  kinematic axis from our dynamical modeling.  Residuals
  (data\,$-$\,model) are shown in panel {\it (e)} on the same velocity
  scale as the velocity and best-fit model.  The final panel shows the
  one-dimensional rotation curve, extracted along the major kinematic
  axis with a pseudo-slit of width 0.5\,$\times$\,FWHM of the seeing
  disk.}
\label{fig:DynamicsExamples}
\end{figure*}

\begin{figure}
  \centerline{\psfig{file=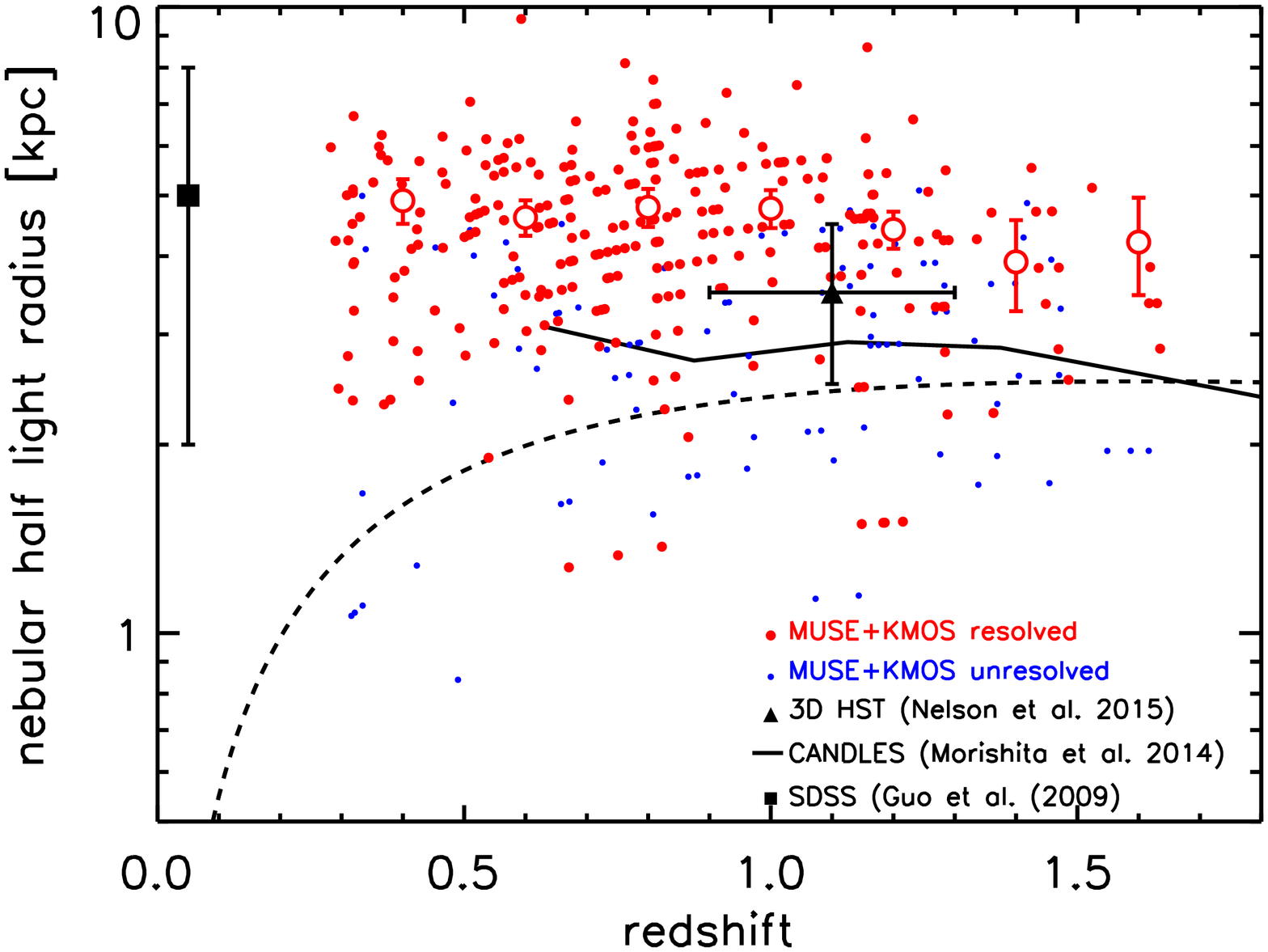,width=3.5in,angle=90}}
  \caption{Evolution of the physical half-light radii with redshift
    for the galaxies in our sample.  We plot the nebular emission line
    sizes in all cases ([O{ \sc ii}] for MUSE or H$\alpha$ for KMOS).
    We plot both the extended (red) and unresolved/compact (blue)
    galaxies individually, but also show the median half-light radii
    in $\Delta z$\,=\,0.2 bins as large filled points with errors
    (these medians include unresolved sources).  We also include
    recent measurements of the nebular emission line half-light radii
    of $z\sim$\,1 galaxies from the \emph{3D-HST} survey
    \citep{Nelson15} and the evolution in the continuum sizes
    (corrected to nebular sizes using the results from
    Fig~\ref{fig:compare_sizes}) from \citep{Morishita14} for galaxies
    in the CANDLES fields.  We also include the size measurements from
    SDSS \citep{Guo09}.  As a guide, the dashed line shows the half
    light radius as a function of redshift for a 0.7$''$ PSF (the
    median seeing of our observations).  This plot shows that the
    nebular emission half-light radii of the galaxies in our sample
    are consistent with similar recent measurements of galaxy sizes
    from \emph{HST} \citep{Nelson15}, and a factor
    $\sim$\,1.5\,$\times$ smaller than late-type galaxies at
    $z$\,=\,0.}
\label{fig:rh_z}
\end{figure}

To create intensity, velocity and velocity dispersion maps for each
galaxy in our MUSE sample, we first extract a 5\,$\times$\,5$''$
``sub-cube'' around each galaxy (this is increased to
7\,$\times$\,7$''$ if the [O{\sc ii}] is very extended) and then fit
the [O{\sc ii}] emission line doublet pixel-by-pixel.  We first
average over 0.6\,$\times$\,0.6$''$ pixels and attempt the fit to the
continuum plus emission lines.  During the fitting procedure, we
account for the increased noise around the sky OH residuals, and also
account for the the spectral resolution (and spectral line spread
function) when deriving the line width.  We only accept the fit if the
improvement over a continuum-only fit is $>$5\,$\sigma$.  If no fit is
achieved, the region size is increased to 0.8\,$\times$\,0.8$''$ and
the fit re-attempted.  In each case, the continuum level, redshift,
line width, and intensity ratio of the 3726.2\,/\,3728.9\AA\, [O{\sc
    ii}] emission line doublet is allowed to vary.  In cases that meet
the signal-to-noise threshold, errors are calculated by perturbing
each parameter in turn, allowing the other parameters to find their
new minimum, until a $\Delta\chi^2$\,=\,1-$\sigma$ is reached.  For
the KMOS observations we follow the same procedure, but fit the
H$\alpha$ and [N{\sc ii}]\,6548,6583 emission lines.  In
Fig.~\ref{fig:DynamicsExamples} we show example images and velocity
fields for the galaxies in our sample (the full sample, along with
their spectra are shown in Appendix~A).  In
Fig.~\ref{fig:DynamicsExamples} the first three panels show the
\emph{HST} image, with ellipses denoting the disk radius and lines
identifying the major morphological and kinematic axis (see
\S~\ref{sec:DynModel}), the MUSE $I$-band continuum image and the
two-dimensional velocity field.  We note that for each galaxy, the
high-resolution image (usually from \emph{HST}) is astrometrically
aligned to the MUSE or KMOS cube by cross correlating the (line free)
continuum image from the cube.

%
%

The ratio of circular velocity (or maximum velocity if the dynamics
are not regular) to line-of-sight velocity dispersion
($V$\,/\,$\sigma$) provides a crude, but common way to classify the
dynamics of galaxies in to rotationally- version dispersion- dominated
systems.  To estimate the maximum circular velocity, $V$, we extract
the velocity profile through the continuum center at a position angle
that maximises the velocity gradient.  We inclination correct this
value using the continuum axis ratio from the broad-band continuum
morphology (see \S~\ref{sec:DynModel}).  For the full sample, we find
a range of maximum velocity gradients from 10 to 540\,km\,s$^{-1}$
(peak-to-peak) with a median of 98\,$\pm$\,5\,km\,s$^{-1}$ and a
quartile range of 48--192\,km\,s$^{-1}$.  To estimate the intrinsic
velocity dispersion, we first remove the effects of beam-smearing (an
effect in which the observed velocity dispersion in a pixel has a
contribution from the intrinsic dispersion and the flux-weighted
velocity gradient across that pixel due to the PSF).  To derive the
intrinsic velocity dispersion, we calculate and subtract the
luminosity weighted velocity gradient across each pixel and then
calculate the average velocity dispersion from the corrected
two-dimensional velocity dispersion map.  In this calculation, we omit
pixels that lie within the central PSF FWHM (typically
$\sim$\,0.6$''$; since this is the region of the galaxy where the
beam-smearing correction is most uncertain).  For our sample, the
average (corrected) line-of-sight velocity dispersion is
$\sigma$\,=\,32\,$\pm$\,4\,km\,s$^{-1}$ (in comparison, the average
velocity dispersion measured from the galaxy integrated
one-dimensional spectrum is $\sigma$\,=\,70\,$\pm$\,5\,km\,s$^{-1}$).
This average intrinsic velocity dispersion at the median redshift of
our sample ($z$\,=\,0.84) is consistent with the average velocity
dispersion seen in a number of other high-redshift samples
\citep[e.g.\ ][]{ForsterSchreiber09,Law09,Gnerucci11,Epinat12,Wisnioski15}.

\begin{figure}
  \centerline{\psfig{file=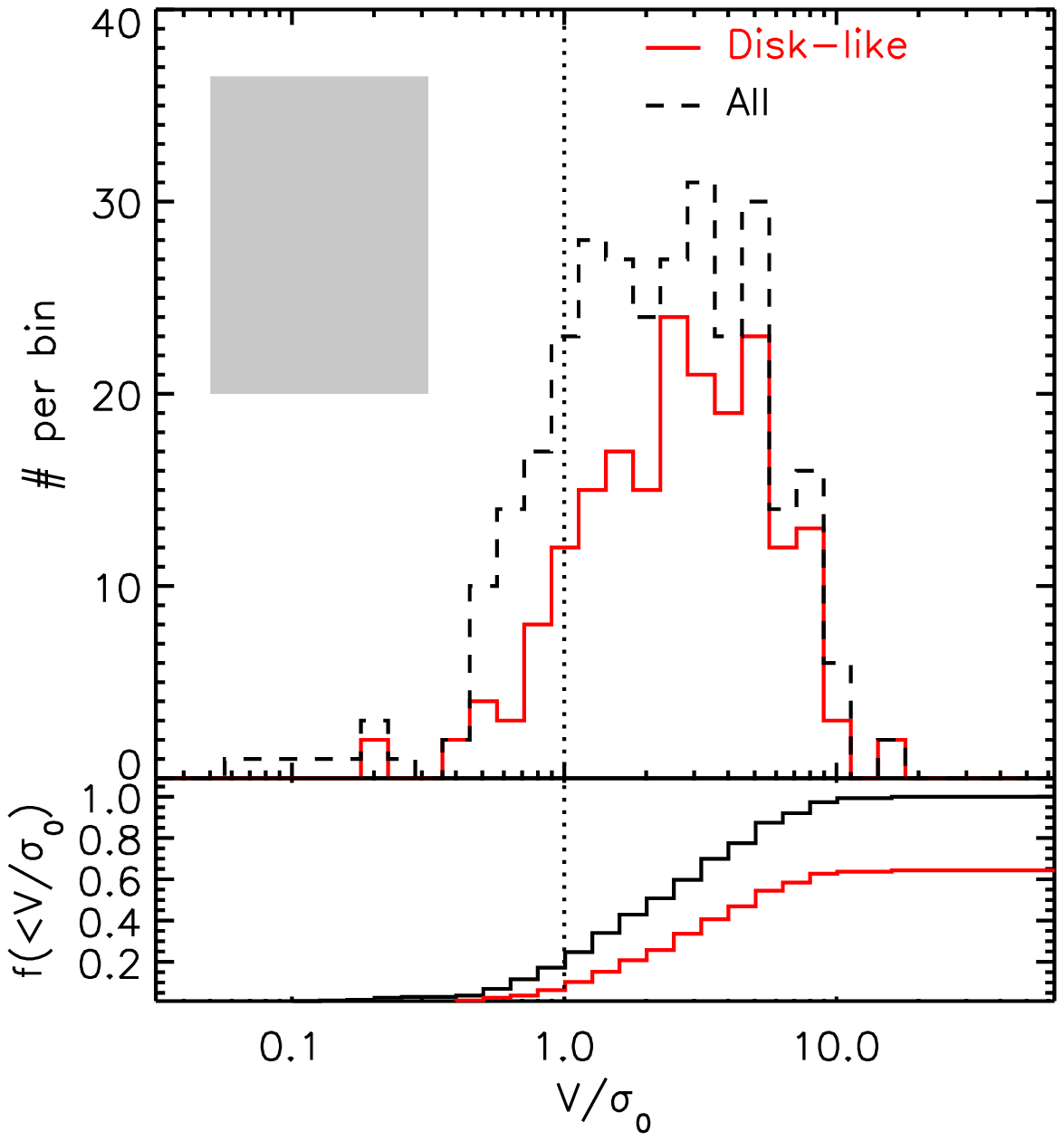,width=3.5in,angle=0}}
  \caption{The ratio of circular velocity to velocity dispersion for
    the galaxies in our sample (V\,/\,$\sigma$), split by their
    classification (the lower panel shows the cumulative
    distribution).  The circular velocity has been inclination
    corrected, and the velocity dispersion has been corrected for
    beam-smearing effects.  The dashed line shows all of the galaxies
    in our sample which are spatially resolved.  The red solid line
    denotes galaxies which are classified as disk-like.  The grey box
    denotes the area occupied by the galaxies that are classified as
    unresolved.  Finally, the dotted line shows a ratio of
    $V$\,/\,$\sigma$\,=\,1.  90\% of the galaxies that are classified
    as disk-like (i.e.\ a spider-line pattern in the velocity field,
    the line-of-sight velocity dispersion peaks near the dynamical
    center of the galaxy and the rotation curve rises smoothly) have
    V\,/\,$\sigma>$\,1, and 67\% have $V$\,/\,$\sigma>$\,2.}
\label{fig:vs_type}
\end{figure}

For the full sample of galaxies in our survey, we derive a median
inclination corrected ratio of $V$\,/\,$\sigma$\,=\,2.2\,$\pm$\,0.2
with a range of $V$\,/\,$\sigma$\,=\,0.1--10 (where we use the limits
on the circular velocities for galaxies classed as unresolved or
irregular\,/\,face-on).  We show the full distribution in
Fig.~\ref{fig:vs_type}.

Although the ratio of $V$\,/\,$\sigma$ provides a means to separate
``rotationally dominated'' galaxies from those that are dispersion
supported, interacting or merging can also be classed as rotationally
supported.  Based on the two-dimensional velocity field, morphology
and velocity dispersion maps, we also provide a classification of each
galaxy in four broad groups (although in the following dynamical
plots, we highlight the galaxies by $V$\,/\,$\sigma$ and their
classification):

\noindent {\it (i)} Rotationally supported: for those galaxies whose
dynamics appear regular (i.e.\ a spider-line pattern in the velocity
field, the line-of-sight velocity dispersion peaks near the dynamical
center of the galaxy and the rotation curve rises smoothly), we
classify as rotationally supported (or ``Disks'').  We further
sub-divide this sample in to two subsets: those galaxies with the
highest-quality rotation curves ($q$\,=\,1; i.e. the rotation curve
appears to flatten or turn over), and those whose rotation curves do
not appear to have asymptoted at the maximum radius determined by the
data ($q$\,=\,2).  This provides an important distinction since for a
number of $q$\,=\,2 cases the asymptotic rotation speed must be
extrapolated (see \S~\ref{sec:VelMeasure}).  The images, spectra,
dynamics and broad-band SEDs for these galaxies are shown in Fig.~A1.

\noindent {\it (ii)} Irregular: A number of galaxies are clearly
resolved beyond the seeing, but display complex velocity fields and
morphologies, and so we classify as ``Irregular''.  In many of these
cases, the morphology appears disturbed (possibly late stage
minor\,/\,major mergers) and\,/\,or we appear to be observing systems
(close-to) face-on (i.e. the system is spatially extended by there is
little\,/\,no velocity structure discernable above the errors).  The
images, spectra, dynamics and broad-band SEDs for these galaxies are
shown in Fig.~A2.

\noindent {\it (iii)} Unresolved: As discussed in
\S~\ref{sec:GalaxySizes}, the nebular emission in a significant
fraction of our sample appear unresolved (or ``compact'') at our
spatial resolution. The images, spectra, dynamics and broad-band SEDs
for these galaxies are shown in Fig.~A3.

\noindent {\it (iv)} Major Mergers: Finally, a number of systems
appear to comprise of two (or more) interacting galaxies on scales
separated by 8--30\,kpc, and we classify these as (early stage) major
mergers. The images, spectra, dynamics and broad-band SEDs for these galaxies are shown in Fig.~A2.

From this broad classification, our [O{\sc ii}] and H$\alpha$ selected
sample comprises 24\,$\pm$\,3\% unresolved systems; 49\,$\pm$\,4\%
rotationally supported systems (27\% and 21\% with $q$\,=\,1 and
$q$\,=\,2 respectively); 22\,$\pm$\,2\% irregular (or face-on) and
$\sim$\,5\,$\pm$\,2\% major mergers.  Our estimate of the ``disk''
fraction in this sample is consistent with other dynamical studies
over a similar redshift range which found that rotationally supported
systems make up $\sim$\,40--70\% of the H$\alpha$- or [O{\sc
    ii}]-selected star-forming population
\citep[e.g.\ ][]{ForsterSchreiber09,Puech08,Epinat12,Sobral13b,Wisnioski15,Stott16,Contini16}.

From this classification, the ``rotationally supported'' systems are
(unsurprisingly) dominated by galaxies with high $V$\,/\,$\sigma$,
with 176/195 (90\%) of the galaxies classed as rotationally
supoprted with V/$\sigma>1$ (and 132\,/\,195 [67\%] with V\,/\,$\sigma>$2).
Concentrating only on those galaxies that are classified as
rotationally supported systems (\S~\ref{sec:ResolvedDynamics}), we
derive $V$\,/\,$\sigma$\,=\,2.9\,$\pm$\,0.2 [3.4\,$\pm$\,0.2 and
  1.9\,$\pm$\,0.2 for the $q$\,=\,1 and $q$\,=\,2 sub-samples
  respectively].  We note that 23\% of the galaxies that are
classified as rotationally supported have $V$\,/\,$\sigma<1$ (21\%
with $q$\,=\,1 and 24\% with $q$\,=\,2).

%
%

%
%

\subsection{Dynamical Modeling}
\label{sec:DynModel}

For each galaxy, we model the
broad-band continuum image and two-dimensional velocity field with a
disk\,+\,halo model.  In addition to the stellar and gaseous disks, the
rotation curves of local spiral galaxies imply the presence of a dark
matter halo, and so the velocity field can be characterized by
$$v^2\,=\,v^2_{\rm d}\,+\,v^2_{\rm h}\,+\,v^2_{\rm HI}$$ where the
subscripts denote the contribution of the baryonic disk
(stars\,+\,H$_2$), dark halo and extended H{\sc i} gas disk respectively.  For
the disk, we assume that the baryonic surface mass density follows an
exponential profile \citep{Freeman70} $$\Sigma_{\rm d}(r) =
{M_{\rm d} \over 2 \pi R_{\rm d}^2} e^{-r/R_{\rm_d}}$$ where $M_{\rm
  d}$ and $R_{\rm d}$ are the disk mass and disk scale length
respectively.  The contribution of this disk to the circular velocity
is:
$$v^2_{\rm D}(x)\,=\,{1\over 2}{G M_{\rm d}\over R_{\rm d}}\,(3.2\,x)^2\,(I_0K_0-I_1K_1)$$
where
$x$\,=\,$R$\,/\,$R_{\rm d}$ and $I_{\rm n}$ and $K_{\rm n}$ are the
modified Bessel functions computed at 1.6\,$x$.
For the dark matter component we assume
$$v_{\rm h}^2(r)\,=\,GM_{\rm  h}(<r)\,/\,r$$ with
$$\rho (r)={\rho_0\, r_0^3 \over (r+r_0)\,(r^2+r_0^2)}$$
\citep{Burkert95,Persic88,Salucci00} where $r_0$ is the core radius
and $\rho_0$ the effective core density.  It follows that
$$M_{\rm
  h}(r)\,=\,4\,M_0\,\left[\ln\left(1+\frac{r}{r_0}\right)\,-\,\tan^{-1}\left(\frac{r}{r_0}\right)\,+\,\frac{1}{2}\,\ln\left(1+\frac{r^2}{r_0^2}\right)\right]$$
with $M_{\rm 0}$\,=\,1.6\,$\rho_0\,r_0^3$ and
{\small $$v^2_H(r)\,={6.4\,G\,\rho_0\,r_0^3\over
    r}\Big\{\ln\Big(1\,+\,\frac{r}{r_0}\Big)\,-\,\tan^{-1}\Big(\frac{r}{r_0}\Big)\,+\,{1\over
    2}\ln\Big[1\,+\,\Big(\frac{r}{r_0}\Big)^2\Big]\Big\}$$ }  This
velocity profile is generic: it allows a distribution with a core of
size $r_0$, converges to the NFW profile \citep{NFW} at large
distances and, for suitable values of $r_0$, it can mimic the NFW or
an isothermal profile over the limited region of the galaxy which is
mapped by the rotation curve.

In luminous local disk galaxies the H{\sc i} disk is the dominant
baryonic component for $r > 3 R_d$.  However, at smaller radii the
H{\sc i} gas disk is negligible, with the dominant component in stars.
Although we can not exclude the possibility that some fraction of
H{\sc i} is distributed within 3\,$R_d$ and so contributes
to the rotation curve, for simplicity, here we assume that the
fraction of H{\sc i} is small and so set $v_{\rm HI}$\,=\,0.

To fit the the dynamical models to the observed images and velocity
fields, we use an MCMC algorithm.  We first use the imaging data to
estimate of the size, position angle and inclination of the galaxy
disk.  Using the highest-resolution image, we fit the galaxy image
with a disk model, treating the [$x_{\rm im}$,$y_{\rm im}$] center,
position angle (PA$_{\rm im})$, disk scale length ($R_{\rm d}$) and
total flux as free parameters.  We then use the best-fit parameter
values from the imaging as the first set of prior inputs to the code
and simultaneously fit the imaging\,+\,velocity field using the model
described above.  For the dynamics, the mass model has five free
parameters: the disk mass ($M_{\rm d}$), radius ($R_{\rm d}$), and
inclination ($i$), the core radius $r_{\rm 0}$, and the central core
density $\rho_{\rm 0}$.  We allow the dynamical center of the disk
([$x_{\rm dyn}$,$y_{\rm dyn}$]) and position angle (PA$_{\rm dyn}$) to
vary, but require that the imaging and dynamical center lie within
1\,kpc (approximately the radius of a bulge at
$z\sim$\,1; \citealt{Bruce14}).  We note also that we allow the
morphological and dynamical major axes to be independent (but see
\S\ref{sec:KinematicMorph_PA}).

To test whether the parameter values returned by the disk modeling
provide a reasonably description of the data, we perform a number of
checks, in particular to test the reliability of recovering the
dynamical center, position angle and disk inclination (since these
propagate directly in to the extraction of the rotation curve and
hence our estimate of the angular momentum).

First, we attempt to recover the parameters from a set of idealized
images and velocity fields constructed from a set of realistic disk
and halo masses, sizes, dynamical centers, inclinations and position
angles.  For each of these models, we construct a datacube from the
velocity field, add noise appropriate for our observations, and then
re-fit the datacube to derive an ``observed'' velocity field.  We then
fit the image and velocity field simultaneously to derive the output
parameters.  Only allowing the inclination to vary (i.e. fixing
[$M_{\rm d}$, $R_{\rm d}$, $\rho_{\rm 0}$, $r_{\rm 0}$, $x_{\rm c}$,
  $y_{\rm c}$, PA] at their input values), we recover the
inclinations, with $i^{\rm in}$\,=\,$i^{\rm out}$\,$\pm$\,2$^{\circ}$.
Allowing a completely unconstrained fit returns inclinations which are
higher than the input values, ($i^{\rm in}$\,/\,$i^{\rm
  out}$\,=\,1.2\,$\pm$\,0.1), the scatter in which can be attributed
to degeneracies with other parameters.  For example, the disk masses
and disk sizes are over-estimated (compared to the input model), with
$M_{\rm d}^{\rm in}$\,/\,$M_{\rm d}^{\rm out}$\,=\,0.86\,$\pm$\,0.12
and $R_{\rm d}^{\rm in}$\,/\,$R_{\rm d}^{\rm
  out}$\,=\,0.81\,$\pm$\,0.05, but the position angle of the major
axis of the galaxy is recovered to within one degree (PA$_{\rm
  in}$\,$-$\,PA$_{\rm out}$\,=\,0.9\,$\pm$\,0.7$^{\circ}$).  For the
purposes of this paper, since we are primarily interested in
identifying the major kinematic axis (the on-sky position angle),
extracting a rotation curve about this axis and correcting for
inclination effects, the results of the dynamical modeling appear as
sufficiently robust that meaningful measurements can be made.

Next, we test whether the inclinations derived from the morphologies
alone are comparable to those derived from a simultaneous fit to the
images and galaxy dynamics.  To obtain an estimate of the inclination,
we use {\sc galfit} \citep{Peng02} to model the morphologies for all
of the galaxies in our sample which have \emph{HST} imaging.  The
ellipticity of the projected image is related to the inclination angle
through cos$^2$\,$i$\,=\,$((b/a)^2-q_0^2)/(1-q_0)^2$ where $a$ and $b$
are the semi-major and semi-minor axis respectively (here $i$ is the
inclination angle of the disk plane to the plane of the sky and
$i$\,=\,0 represents an edge-on galaxy).  The value of $q_0$ (which
accounts for the fact that the disks are not thin) depends on galaxy
type, but is typically in the range $q_0$\,=\,0.13--0.20 for
rotationally supported galaxies at $z\sim$\,0, and so we adopt
$q_0$\,=\,0.13.  We first construct the point-spread function for each
\emph{HST} field using non-saturated stars in the field of view, and
then run {\sc galfit} with Sersic index allowed to vary from
$n$\,=\,0.5--7 and free centers and effective radii.
For galaxies whose dynamics resemble rotating systems (such that a
reasonable estimate of the inclination can be derived) the inclination
derived from the morphology is strongly correlated with that inferred
from the dynamics, with a median offset of just $\Delta i$\,=\,4$^{\circ}$
with a spread of $\sigma_i$\,=\,12$^{\circ}$.

The images, velocity fields, best-fit kinematic maps and velocity
residuals for each galaxy in our sample are shown in Fig.~A1--A3, and the
best-fit parameters given in Table~A1.  Here, the errors reflect the
range of acceptable models from {\it all} of the models attempted.
All galaxies show small-scale deviations from the best-fit model, as
indicated by the typical r.m.s,
$<$\,data\,$-$\,model\,$>$\,=\,28\,$\pm$\,5\,km\,s$^{-1}$.  These
offsets could be caused by the effects of gravitational instability,
or simply be due to the un-relaxed dynamical state indicated by the high
velocity dispersions in many cases.
The goodness of fit and small-scale deviations from the best-fit
models are similar to those seen in other dynamical surveys of
galaxies at similar redshifts, such as KMOS$^{\rm 3D}$ and KROSS
(\citealt{Wisnioski15,Stott16}) where rotational support is also seen
in the majority of the galaxies (and with r.m.s of
10--80\,km\,s$^{-1}$ between the velocity field and best-fit disk
models).

\subsection{Kinematic versus Morphological Position Angle}
\label{sec:KinematicMorph_PA}

One of the free parameters during the modeling is the offset between
the major morphological axis and the major dynamical axis.  The
distribution of misalignments may be attributed to physical
differences between the morphology of the stars and gas, extinction
differences between the rest-frame UV\,/\,optical and H$\alpha$,
sub-structure (clumps, spiral arms and bars) or simply measurement
errors when galaxies are almost face on.
Following \citet{Franx91} \citep[see also][]{Wisnioski15}, we define
the misalignment parameter, $\Phi$, such that
sin\,$\Phi$\,=\,$|$sin(PA$_{\rm phot}$\,$-$\,PA$_{\rm dyn}$)$|$ where
$\Phi$ ranges from 0--90$^{\circ}$.  For all of the galaxies in our
sample whose dynamics resemble rotationally supported systems, we
derive a median ``misalignment'' of
$\Phi$\,=\,9.5\,$\pm$\,0.5$^{\circ}$
($\Phi$\,=\,10.1\,$\pm$\,0.8$^{\circ}$ and 8.6\,$\pm$\,0.9$^{\circ}$
for $q$\,=\,1 and $q$\,=\,2 sub-samples respectively).
In all of the following sections, when extracting rotation curves (or
velocities from the two-dimensional velocity field), we use the
position angle returned from the dynamical modeling, but note that
using the morphological position angle instead would reduce the
peak-to-peak velocity by $\lsim$\,5\%, although this would have no
qualitative effect on our final conclusions.

\subsection{Velocity Measurements}
\label{sec:VelMeasure}

To investigate the various velocity--stellar mass and angular momentum
scaling relations, we require determination of the circular velocity.
For this analysis, we use the best-fit dynamical models for each
galaxy to make a number of velocity measurements.  We measure the
velocity at the ``optical radius'', $V$(3\,$R_{\rm d}$)
\citep{Salucci00} (where the half light- and disk- radius are related
by $r_{\rm 1/2}$\,=\,1.68\,R$_{\rm d}$).  Although we are using the
dynamical models to derive the velocities (to reduce errors in
interpolating the rotation curve data points), we note that the
average velocity offset between the data and model for the
rotationally supported systems at $r_{\rm 1/2}$ is small, $\Delta
V$\,=\,2.1\,$\pm$\,0.5\,km\,s$^{-1}$ and $\Delta
V$\,=\,2.4\,$\pm$\,1.2\,km\,s$^{-1}$ at 3\,$R_{\rm d}$.  In 30\% of
the cases, the velocities at 3\,$R_{d}$ are extrapolated beyond the
extent of the observable rotation curve, although the difference
between the velocity of the last data point on the rotation curve and
the velocity at 3\,$R_{\rm d}$ in this sub-sample is only $\Delta
v$\,=\,2\,$\pm$\,1\,km\,s$^{-1}$ on average.

\subsection{Angular Momentum}

With measurements of (inclination corrected) circular velocity, size
and stellar mass of the galaxies in our sample, we are in a position
to combine these results and so measure the specific angular momentum
of the galaxies (measuring the specific angular momentum removes the
implicit scaling between $J$ and mass).  The specific angular momentum
is given by
\begin{equation}
  j_\star=\frac{J}{M_\star}\,=\,\frac{\int_{\bf r}({\bf r}\times {\bf \bar{v}})\rho_\star\, d^3{\bf r}}{\int_{\bf r}\rho_\star\,d^3{\bf r}}
  \label{eqn:jt_3D}
\end{equation}
where $r$ and $\bar{v}(r)$ are the position and mean-velocity vectors
(with respect to the center of mass of the galaxy) and $\rho(r)$ is
the three dimensional density of the stars and gas.

To enable us to compare our results directly with similar measurements
at $z\sim$\,0, we take the same approximate estimator for specific
angular momentum as used in \citet{Romanowsky12} (although see
\citealt{Burkert16} for a more detailed treatment of angular momentum
at high-redshift).  In the local samples of \citet{Romanowsky12} (see
also \citet{Obreschkow15}), the scaling between specific angular
momentum, rotational velocity and disk size for various morphological
types is given by
\begin{equation}
  j_n = k_n\,C_i\,v_s R_{1/2}
  \label{eqn:jvr}
\end{equation}
where $v_{\rm s}$ is the rotation velocity at 2$\times$ the half-light
radii ($R_{\rm 1/2}$) (which corresponds to $\simeq3\,R_{\rm D}$ for
an exponential disk), $C_{\rm i}$\,=\,sin$^{-1}\theta_{\rm im}$ is the
deprojection correction factor (see \citealt{Romanowsky12}) and
$k_{\rm n}$ depends on the Sersic index ($n$) of the galaxy which can
be approximated as
\begin{equation}
  k_n\,=\,1.15\,+\,0.029\,n\,+\,0.062\,n^2
  \label{eqn:kn}
\end{equation}
For the galaxies with $HST$ images, we run {\sc galfit} to estimate
the sersic index for the longest-wavelength image available and derive
a median sersic index of $n$\,=\,0.8$\pm$0.2, with 90\% of the sample
having $n<2.5$, and therefore we adopt $j_\star$\,=\,$j_{n=1}$, which
is applicable for exponential disks.  Adopting a sersic index of
$n$\,=\,2 would result in a $\sim$\,20\% difference in $j_\star$.
To infer the circular velocity, we measure the velocity from the
rotation curve at 3\,$R_d$; \citealt{Romanowsky12}).  We report all of
our measurements in Table~A1.

\begin{figure*}
  \centerline{\psfig{file=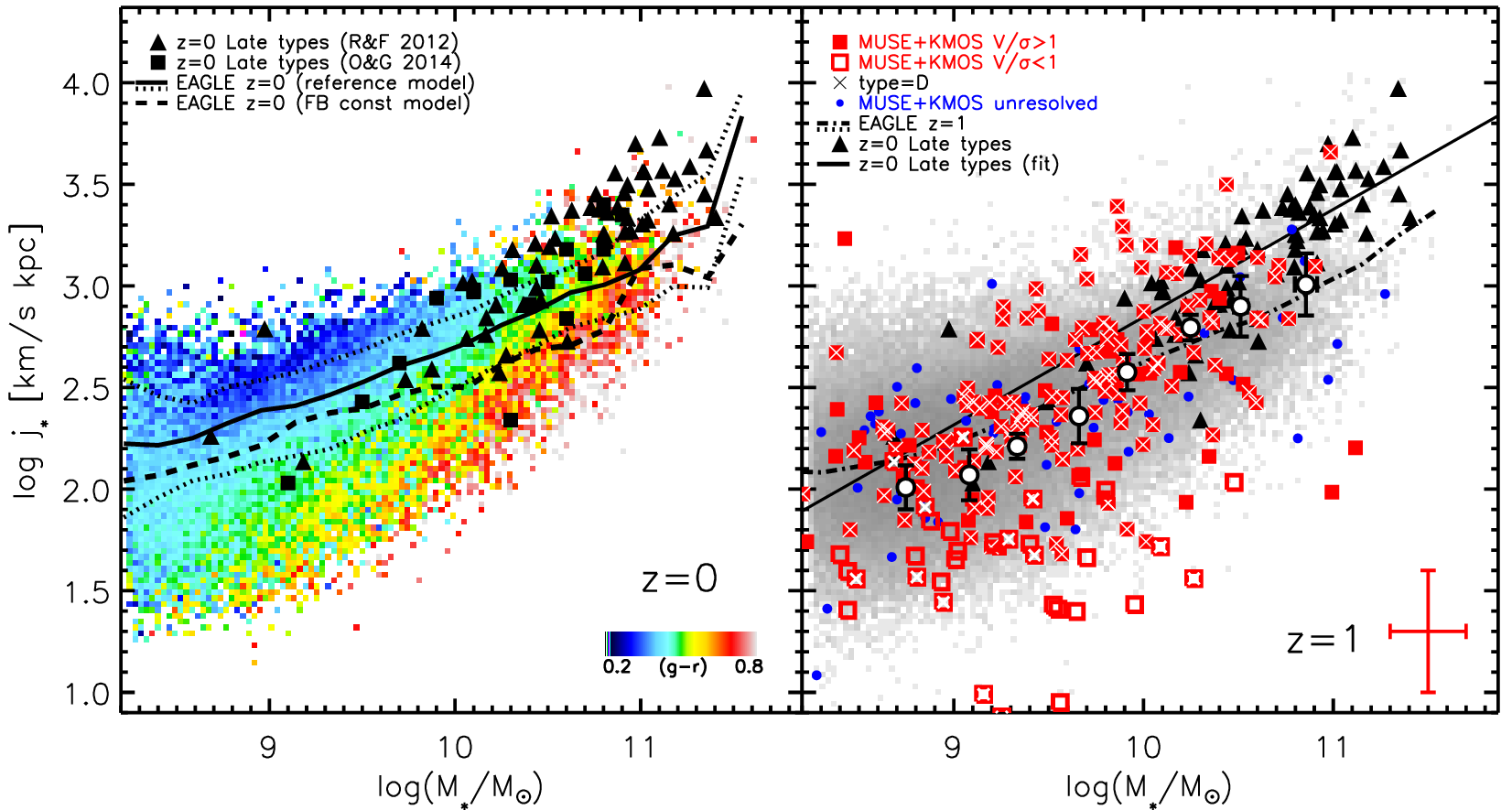,width=7in,angle=90}}
  \caption{{\it Left:} Specific angular momentum
    ($j_\star$\,=\,$J$\,/\,M$_\star$) of late- and early- type
    galaxies at $z$\,=\,0 from \citet{Romanowsky12} and
    \citet{Obreschkow14} (R\&F\,2012 and O\&G 2014 respectively), both
    of which follow a scaling of $j_\star\propto$\,M$_\star^{2/3}$.
    We also show the specific angular momentum of galaxies at
    $z$\,=\,0 from the {\sc eagle} simulation (reference model) with
    the colour scale set by the rest-frame $g-r$ colours of the
    galaxies.  The solid line shows the median (and dotted lines
    denote the 68\% distribution width) of the {\sc eagle} galaxies.
    For comparison with other {\sc eagle} models, we also include the
    evolution of $j_\star$--M$_\star$ from the ``constant feedback''
    {\it FBconst} model (dashed line).  {\it Right:} The specific
    angular momentum for the high-redshift galaxies in our MUSE and
    KMOS sample.  We split the high-redshift sample in to those
    galaxies with the best sampled dynamics\,/\,rotation curves (which
    we denote $q$\,=\,1) and those with less well constrained rotation
    curves ($q$\,=\,2).  In the lower right corner we show the typical
    error bar, estimated using a combination of errors on the stellar
    mass, and uncertainties in the inclination and circular velocity
    measurement.  We also include on the plot the unresolved galaxies
    from our sample using the limits on their sizes and velocity
    dispersions (the latter to provide an estimate of the upper limit
    on $v_{\rm c}$).  The median specific angular momentum (and
    bootstrap error) in bins of log$_{\rm 10}$(M$_\star$)\,=\,0.3\,dex
    is also shown.  The grey-scale shows the predicted distribution at
    $z\sim$\,1 from the {\sc eagle} simulation and we plot the median
    specific angular momentum in bins of stellar mass as well as the
    {\sc eagle} $z$\,=\,0 model from the left-hand panel.  Although
    there is considerable scatter in the high-redshift galaxy sample,
    at $z\sim$\,1, there are very few high stellar mass galaxies with
    specific angular momentum as large as comparably massive local
    spirals, suggesting that most of the accretion of high angular
    momentum material must occur below $z\sim$\,1.}
  \label{fig:Jz}
\end{figure*}

In Fig.~\ref{fig:Jz} we plot the specific angular momentum versus
stellar mass for the high-redshift galaxies in our sample and compare
to observations of spiral galaxies at $z$\,=\,0
\citep{Romanowsky12,Obreschkow14}.  We split the high-redshift sample
in to those galaxies with the best sampled dynamics\,/\,rotation
curves ($q$\,=\,1) and those with less well constrained dynamics
($q$\,=\,2).  To ensure we are not biased towards large\,/\,resolved
galaxies in the high-redshift sample, we also include the unresolved
galaxies, but approximate their maximum specific angular momentum by
$j_\star$\,=\,1.3\,$r_{\rm 1/2}$\,$\sigma$ (where $\sigma$ is the
velocity dispersion measured from the collapsed, one-dimensional
spectrum and is assumed to provide an upper limit on the circular
velocity.  The pre-factor of 1.3 is derived assuming a Sersic index of
$n$\,=\,1--2; \citealt{Romanowsky12}).  We note that three of our survey
fields (PKS1614$-$9323, Q2059$-$360 and Q0956+122) do not have
extensive multi-wavelength imaging required to derive stellar masses
and so do not include these galaxies on the plot.

%
%

\subsection{{\sc eagle} Galaxy Formation Model}
Before discussing the results from Fig.~\ref{fig:Jz}, we first need to
test whether there may be any observational selection biases that may
affect our conclusions.  To achieve this, and aid the interpretation
of our results, we exploit the hydro-dynamic {\sc eagle}
simulation.  We briefly discuss this simulation here, but refer the
reader for \citep[][and references therein]{Schaye15} for a details.
The Evolution and Assembly of GaLaxies and their Environments {\sc
  (eagle)} simulations follows the evolution of dark matter, gas,
stars and black-holes in cosmological (10$^6$\,Mpc$^3$) volumes
\citep{Schaye15,Crain15}.  The {\sc eagle} reference model is
particularly useful as it provide a resonable match to the present-day
galaxy stellar mass function, the amplitude of the galaxy-central
black hole mass relation, and matches the $z\sim$\,0 galaxy sizes and
the colour--magnitude relations.  With a reasonable match to the
properties of the $z\sim$\,0 galaxy population, {\sc eagle} provides a
useful tool for searching for, and understanding, any observational
biases in our sample and also for interpreting our results.

Lagos et al.\ (2016) show that the redshift evolution of the specific
angular momentum of galaxies in the {\sc eagle} simulation depends
sensitively on mass and star formation rate cuts applied.  For
example, in the model, massive galaxies which are classified as
``passive'' around $z\sim$\,0.8 (those well below the
``main-sequene'') show little\,/\,no evolution in specific angular
momentum from $z\sim$\,0.8 to $z$\,=\,0, whilst ``active''
star-forming galaxies (i.e.\ on or above the ``main-sequence'') can
increase their specific angular momentum\footnotemark as rapidly as
$j_\star$\,/\,M$_\star^{2/3}\propto(1+z)^{3/2}$.
In principle, these predictions can be tested by observations.
\footnotetext{We note that in the angular momentum comparisons below,
  quantitatively similar results have been obtained from the Illustris
  simulation \citep{Genel15}}.

From the {\sc eagle} model, the most direct method for calculating angular
momentum galaxies is to sum the angular momentum of each star particle
that is associated with a galaxy ($J_{\rm
  p}$\,=\,$\sum_i$\,m$_i$\,${\bf r_i}\times$\,${\bf v_i}$).  However, this does
not necessarily provide a direct comparison with the observations
data, where the angular momentum is derived from the rotation curve
and a measured galaxy sizes.
To ensure a fair comparison between the observations and model can be
made, we first calibrate the particle data in the {\sc eagle} galaxies
with their rotation curves.  \citet{Schaller15} extract rotation
curves for {\sc eagle} galaxies and show that over the radial range
where the galaxies are well resolved, their rotation curves are in
good agreement with those expected for observed galaxies of similar
mass and bulge-to-disk ratio.  We therefore select a subset of 5\,000
galaxies at $z\sim$\,0 from the {\sc eagle} simulation that have
stellar masses between M$_\star$\,=\,10$^8$--10$^{11.5}$\,M$_{\odot}$
and star formation rates of SFR\,=\,0.1--50\,M$_\odot$\,yr$^{-1}$
(i.e.\ reasonably well matched to the mass and star formation rate
range of our observational sample) and derive their rotation curves.
In this calculation, we adopt the minimum of their gravitational
potential as the galaxy center.  We measure their stellar half mass
radii ($r_{\rm 1/2,\star}$), and the circular velocity from the
rotation curve at 3\,R$_{\rm d}$ and then compute the angular momentum
from the rotation curve ($J_{\rm RC}$\,=\,M$_\star$\,$r_{\rm
  1/2,\star}$\,V(3\,R$_{\rm d}$)), and compare this to the angular
momentum derived from the particle data ($J_{\rm P}$).  The angular
momentum of the {\sc eagle} galaxies\footnotemark\footnotetext{We note
  that Lagos et al.\ (2016) show that in {\sc eagle} the value of
  $J_\star$ and the scaling between $J_\star$ and stellar mass is
  insensitive to whether an aperture of 5\,$r_{\rm 50}$ or $r_{\rm
    total}$ is used.} measured from the particular data ($J_{\rm P}$)
broadly agrees with that estimated from the rotation curves ($J_{\rm
  RC}$), although fitting the data over the full range of $J$, we
measure a sub-linear relation of log$_{\rm 10}$($J_{\rm
  RC}$)\,=\,(0.87\,$\pm$\,0.10)\,log$_{\rm 10}$($J_{\rm
  P}$)\,+\,1.75\,$\pm$\,0.20.  Although only a small effect, this
sub-linear offset occurs due to two factors.  First, the sizes of the
low-mass galaxies become comparable to the $\sim$\,1\,kpc
gravitational softening length of the simulation; and second, at lower
stellar masses, the random motions of the stars have a larger
contribution to the total dynamical support.  Nevertheless, in all of
the remaining sections (and to be consistent with the observational
data) we first calculate the ``particle'' angular momentum of {\sc eagle}
galaxies and then convert these to the ``rotation-curve'' angular
momentum.

To test how well the {\sc eagle} model reproduces the observed
mass--specific angular momentum sequence at $z$\,=\,0, in
Fig.~\ref{fig:Jz} we plot the specific angular momentum
($j_\star$\,=\,$J$\,/\,$M_\star$) of $\sim$\,50 late- type galaxies
from the observational study of \citet{Romanowsky12} and also include
the observations of 16 nearby spirals from the The H{\sc i} Nearby
Galaxy Survey (THINGS; \citealt{Walter08}) as discussed in
\citet{Obreschkow14}.  As discussed in \S1, these local disks follow a
correlation of $j_{\star}\propto $\,M$_{\star}^{2/3}$ with a scatter
of $\sigma_{\rm log\,j}\sim$\,0.2\,dex.  We overlay the specific
angular momentum of galaxies at $z$\,=\,0 from the the {\sc eagle}
simulation, colour coded by their rest-frame $(g-r)$ colour
\citep{Trayford16}.  This highlights that the {\sc eagle} model
provides a reasonable match to the $z$\,=\,0 scaling in
$j_\star\propto$\,M$_\star^{2/3}$ in both normalisation and scatter.
Furthermore, the colour-coding highlights that, at fixed stellar mass,
the blue star-forming galaxies (late-types) have higher angular
momentum compared than those with redder (early-type) colours.  A
similar conclusion was reported by \citet{Zavala16} who separated
galaxies in {\sc eagle} in to early versus late types using their
stellar orbits, identifying the same scaling between specific angular
momentum and stellar mass for the late-types.  Lagos et al.\ (2016)
also extend the analysis to investigate other morphological proxies
such as spin, gas fraction, ($u-r$) colour, concentration and stellar
age and in all cases, the results indicate that galaxies that have low
specific angular momentum (at fixed stellar mass) are gas poor, red
galaxies with higher stellar concentration and older mass-weighted
ages.

In Fig.~\ref{fig:Jz} we also show the predicted scaling between
stellar mass and specific angular momentum from {\sc eagle} at
$z$\,=\,1 after applying our mass and star formation rate limits to
the galaxies in the model.  This shows that {\sc eagle} predicts the
same scaling between specific angular momentum and stellar mass at
$z$\,=\,0 and $z$\,=\,1 with $j_\star\propto$\,M$_\star^{2/3}$, with a
change in normalisation such that galaxies at $z\sim$\,1 (at fixed
stellar mass) have systematically lower specific angular momentum by
$\sim$\,0.2\,dex than those at $z\sim$\,0.  We will return to this
comparison in \S~\ref{sec:discussion}.

Before discussing the high-redshift data, we note that one of the
goals of the {\sc eagle} simulation is to test sub-grid recipes for
star-formation and feedback.  The sub-grid recipes in the {\sc eagle}
``reference model'' are calibrated to match the stellar mass function
at $z$\,=\,0, but this model is not unique.  For example, in the
reference model the energy from star-formation is coupled to the ISM
according to the local gas density and metallicity.  This density
dependence has the effect that outflows are able to preferentially
expel material from centers of galaxies, where the gas has low angular
momentum.  However, as discussed by \citet{Crain15}, in other {\sc
  eagle} models that also match the $z$\,=\,0 stellar mass function,
the energetics of the outflows are coupled to the ISM in different
ways, with implications for the angular momentum.  For example, in the
{\it FBconst} model, the energy from star formation is distributed
evenly in to the surrounding ISM, irrespective of local density and
metallicity.  Since this model also matches the $z$\,=\,0 stellar mass
function, and so it is instructive to compare the angular momentum of
the galaxies in this model compared to the reference model.  In
Fig.~\ref{fig:Jz} we also overlay the $z$\,=\,0 relation between the
specific angular momentum ($j_\star$) and stellar mass (M$_\star$) in
the {\sc eagle} {\it FBconst} model.  For stellar masses
M$_\star\gsim$\,10$^{10}$\,M$_\odot$, the specific angular momentum of
galaxies are a factor $\sim$\,2 lower than those in the reference
model.  Since there is no dependence on outflow energetics with local
density, this is a consequence of removing less low angular momentum
material from the disks, which produces galaxies with specific angular
momentum two times smaller than those in the reference model
\citep[][]{Crain15,Furlong15}.  This highlights how observational
constraints on the galaxy angular momentum can play a role in testing
the sub-grid recipes used in numerical simulations.

\subsection{Disk stability}
\label{sec:Disk_V_Sigma}

In \S~\ref{sec:discussion} we will investigate how the specific
angular momentum is related to the galaxy morphologies.  The ``disk
stability'' is intimitely related to the galaxy morphologies, and so
it is instructive to provide a crude (galaxy integrated) measurement
to aid the interpretation of these results.  To define the disk
stability, we use the Toomre parameter \citep{Toomre64}.  In rotating
disk of gas and stars, perturbations smaller than critical wavelength
($\lambda_{\rm max}$) are stabilised against gravity by velocity
dispersion whilst those larger than $\lambda_{\rm min}$ are stabilised
by centrifugal force.  The Toomre parameter is defined by
$Q$\,=\,$\lambda_{\rm min}$\,/\,$\lambda_{\rm max}$, but can also be
expressed as $Q=\sigma\kappa$\,/\,($\pi G\Sigma_{\rm gas}$) where
$\sigma$ is the radial velocity dispersion, $\Sigma$ is the gas
surface density and $\kappa$ is the epicylic frequency.  If $Q<1$,
instabilities can develop on scales larger than the Jeans length and
smaller than the maximal stability scale set by differential rotation.
If $Q>1$, then the differential rotation is sufficiently large to
prevent large scale collapse and no instabilities can develop.

\begin{figure}
  \centerline{\psfig{file=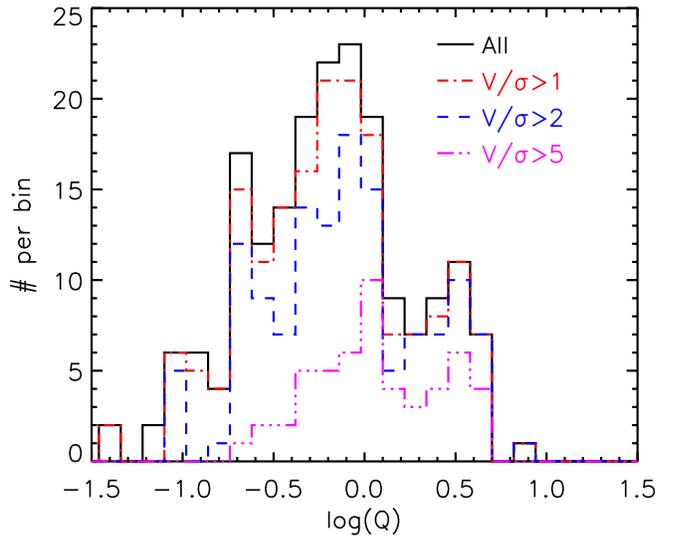,width=3.4in,angle=0}}
  \caption{The distribution of Toomre Q for all galaxies in our sample
    that are classed as rotationally supported.  We also sub-divide
    the sample by the ratio of rotational velocity to velocity
    dispersion (V\,/\,$\sigma$), with V\,/\,$\sigma>$\,1, 2 and 5.
    The full range of Q for the whole sample is $Q$\,=\,0.08--5.6, but
    with increasing V\,/\,$\sigma$, the median Toomre Q also increases
    to Q\,=\,0.80\,$\pm$\,0.10, Q\,=\,0.80\,$\pm$\,0.10,
    Q\,=\,0.90\,$\pm$\,0.08 and Q\,=\,1.30\,$\pm$\,0.16 for the full
    sample to V\,/\,$\sigma>$\,1, 2 and 5 respectively. }
  \label{fig:Q_type}
\end{figure}

To estimate the Toomre $Q$ of each galaxy in our sample, we first
estimate the gas surface density from the redenning corrected star
formation surface density (adopting the total star formation rate
within 2\,$r_{\rm 1/2}$ from \S~\ref{sec:SEDs}) and use the Kennicutt
Schmidt relation \citep{KS98} to infer $\Sigma_{\rm gas}$.  To
estimate the epicyclic frequency of the disk ($\kappa$) we adopt the
(inclination corrected) rotational velocity at 3\,$R_{\rm d}$.  We
also calculate the (beam-smearing corrected) velocity dispersion to
measure $\sigma$.  For the galaxies in our sample that are classified
as rotationally supported, we derive a median Toomre $Q$ of
$Q$\,=\,0.80\,$\pm$\,0.10 (with a full range of $Q$\,=\,0.08--5.6).
On average, these galaxies therefore have disks that are consistent
with being marginally stable.  This is not a surprising result for a
high-redshift [O{\sc ii}] (i.e.\ star formation)-selected sample.  For example,
\citep{Hopkins12b} show that due to feedback from stellar winds,
star-forming galaxies should be driven to the marginally stable
threshold, in particular at high-redshift where the galaxies have high
gas-fractions.  In Fig.~\ref{fig:Q_type} we show the distribution of
Toomre $Q$, split by $V$\,/\,$\sigma$.  Although there are degeneracies
between $Q$ and $V$\,/\,$\sigma$, all of the sub-samples
($V$\,/\,$\sigma>$1, 2, and 5) span the full range in $Q$, although
the median Toomre $Q$ increases with V\,/\,$\sigma$ with
Q\,=\,0.80\,$\pm$\,0.10, Q\,=\,0.90\,$\pm$\,0.08 and
Q\,=\,1.30\,$\pm$\,0.16 for V\,/\,$\sigma>$\,1, 2 and 5 respectively.
We will return to a discussion of this when comparing to the
broad-band morphologies in \S~\ref{sec:discussion}.

Nevertheless, this observable provides a crude, but common way to
classify the stability of the gas in a disk and this will be important
in comparison with the angular momentum.  For example, in local
galaxies Cortese et al. (2016) (using SAMI) and Lagos et al.\ (2016)
(using the {\sc eagle} galaxy formation model) show that the disk
stability and galaxy spin, $\lambda_{\rm R}$ \citep[as defined
  in][]{Emsellem07} are strongly correlated with $V$\,/\,$\sigma$ and
define a continious sequence in the specific angular momentum--stellar
mass plane, where galaxies with high specific angular momentum are the
most stable with high $V$\,/\,$\sigma$ and $\lambda_{\rm R}$.
Moreover, Stevens et al.\ (2016) \citep[see also][]{Obreschkow15}
suggest that specific angular momentum plays a defining role in
defining the disk stability.  We will return this in
\S~\ref{sec:discussion}.

\section{Discussion}
\label{sec:discussion}

Observations of the sizes and rotational velocities of local spiral
galaxies have suggested that $\sim$\,50\% of the initial specific
angular momentum of the baryons within dark matter halos must be lost
due to viscous angular momentum redistribution and selective gas
losses which occur as the galaxy forms and evolves.

In Fig.~\ref{fig:Jz} we plot the specific angular momentum versus
stellar mass for the high-redshift galaxies in our MUSE and KMOS
sample.  In this figure, we split the sample by their dynamics
according to their ratio of $V$\,/\,$\sigma$ (although we also
highlight the galaxies whose dynamics most obviously display
rotational support).  We include the unresolved galaxies from our
sample using the limits on their sizes and velocity dispersions (the
latter to provide an estimate of the upper limit on $V_{\rm c}$).  In
this figure, we also include the distribution (and median+scatter) at
$z$\,=\,0 and $z\sim$\,1 from the {\sc eagle} simulation.

Since there is considerable scatter in the data we bin the specific
angular momentum in stellar mass bins (using bins with d\,log$_{\rm
  10}$(M$_\star$)\,=\,0.3\,dex) and overlay the median (and scatter in
the distribution) in Fig.~\ref{fig:Jz}.  Up to a stellar mass of
$\sim$\,10$^{10.5}$\,M$_\odot$, the high-redshift galaxies follow a
similar scaling between stellar mass and specific angular momentum as
seen in local galaxies \citep[see also][]{Contini16}.  Fitting the
data over the stellar mass range
M$_\star$\,=\,10$^{8.5}$--10$^{11.5}$\,M$_\odot$, we derive a scaling
of $j_\star\propto$\,M$_\star^{\rm q}$ with $q$\,=\,0.6\,$\pm$\,0.1.
Although the scaling $j_\star\propto$\,M$_\star^{2/3}$ is generally
seen in local galaxies, when galaxies are split by morphological type,
the power-law index varies between $q$\,=\,0.7--1 (e.g.\ Cortese et
al.\ 2016).  However, the biggest difference between $z$\,=\,0 and
$z$\,=\,1 is above a stellar mass of
$M_\star\sim$\,10$^{10.5}$\,M$_{\odot}$, where the specific angular
momentum of galaxies at $z\sim$\,1 is 2.5\,$\pm$\,0.5$\times$ lower
than for comparably massive spiral galaxies at $z\sim$\,0, and there
are no galaxies in our observation sample with specific angular
momentum as high as those of local spirals.

First we note that this offset (and lack of galaxies with high
specific angular momentum) does not appear to be driven by volume or
selection effects which result in our observations missing high
stellar mass, high $j_\star$ galaxies.  For example, although the
local galaxy sample from \citet{Romanowsky12} sample is dominated by
local (D\,$<$\,180Mpc) high-mass, edge on spiral disks, the space
density of star-forming galaxies with stellar mass
$>$10$^{11}$\,M$_\odot$ at $z\sim$\,1 is
$\sim$\,1.6\,$\times$\,10$^{-3}$\,Mpc$^{-3}$ \citep{Bundy05}.  The
volume probed by the MUSE and KMOS observations is
$\sim$\,1.5\,$\times$\,10$^4$\,Mpc$^3$ (comoving) between
$z$\,=\,0.4--1.2 and we expect $\sim$\,23\,$\pm$\,4 such galaxies in
our sample above this mass (and we detect 20).  Thus, we do not appear
to be missing a significant population of massive galaxies from our
sample.  At $z\sim$\,1, we are also sensitive to star formation rates
as low as $\sim$\,4\,M$_{\odot}$\,yr$^{-1}$ (given our typical surface
brightness limits and adopting a median reddening of $A_{\rm
  V}$\,=\,0.5).  This is below the so-called ``main-sequence'' at this
redshift since the star formation rate for a ``main-sequence'' galaxy
with M$_\star$\,=\,10$^{11}$\,M$_{\odot}$ at $z$\,=\,1 is
100\,M$_{\odot}$\,yr$^{-1}$ \citep{Wuyts13}.

What physical processes are likely to affect the specific angular
momentum of baryonic disks at high-redshift (particularly those in
galaxies with high stellar masses)?  Due to cosmic expansion, a
generic prediction of $\Lambda$CDM is that the relation between the
mass and angular momentum of dark matter halos changes with time.  In
a simple, spherically symmetric halo the specific angular momentum,
$j_{\rm h}$\,=\,$J_{\rm h}$\,/\,M$_{\rm h}$ should scale as $j_{\rm
  h}$\,=\,M$_{\rm h}^{2/3}(1+z)^{-1/2}$ \citep[e.g.\ ][]{Obreschkow15}
and if the ratio of the stellar-to-halo mass is independent of
redshift, then the specific angular momentum of the baryons should
scale as $j_\star\propto M_\star^{2/3}(1+z)^{-1/2}$.  At $z\sim$\,1,
this simple model predicts that the specific angular momentum of disks
should be $\sqrt{2}$ lower than at $z$\,=\,0.

However, this 'closed-box' model does not account for gas inflows or
outflows, and cosmologically based models have suggested redshift
evolution in $j_\star$\,/\,M$^{2/3}$ can evolve as rapidly as
(1+$z$)$^{-3/2}$ from $z\sim$\,1 to $z$\,=\,0 (although this redshift
evolution is sensitive to the mass and star formation rate limits
applied to the selection of the galaxies; e.g.\ Lagos et al.\ 2016).
For example, applying our mass and star-formation rate limits to
galaxies in the {\sc eagle} model, galaxies at $z\sim$\,1 are
predicted to have specific angular momentum which is 0.2\,dex lower
(or a factor $\sim$\,1.6$\times$) lower than comparably massive
galaxies at $z$\,=\,0, although the most massive spirals at $z$\,=\,0
have specific angular momentum which is $\sim$\,3 times larger than
any galaxies in our high-redshift sample.

The specific angular momentum of a galaxy can be increased or
decreased depending on the evolution of the dark halo, the angular
momentum and impact parameter of accreting material from the
inter-galactic medium, and how the star-forming regions evolve within
the ISM.  For example, if the impact parameter of accreting material
is comparable to the disk radius (as suggested in some models;
e.g.\ \citealt{Dekel09b}), then the streams gradually increase the
specific angular momentum of the disk with decreasing redshift as the
gas accretes on to the outer disk.  The specific angular momentum can
be further increased if the massive, star-forming regions (clumps)
that form within the ISM torque and migrate inwards (since angular
momentum is transferred outwards).  However, galaxy average specific
angular momentum can also be decreased if outflows (associated with
individual clumps) drive gas out of the disk, and
outflows with mass loading factors $\gg$1 associated with
individual star forming regions (clumps) have been observed in a
number of high-redshift galaxies \citep[e.g.\ ][]{Genzel11,Newman12}.

\begin{figure*}
  \centerline{\psfig{file=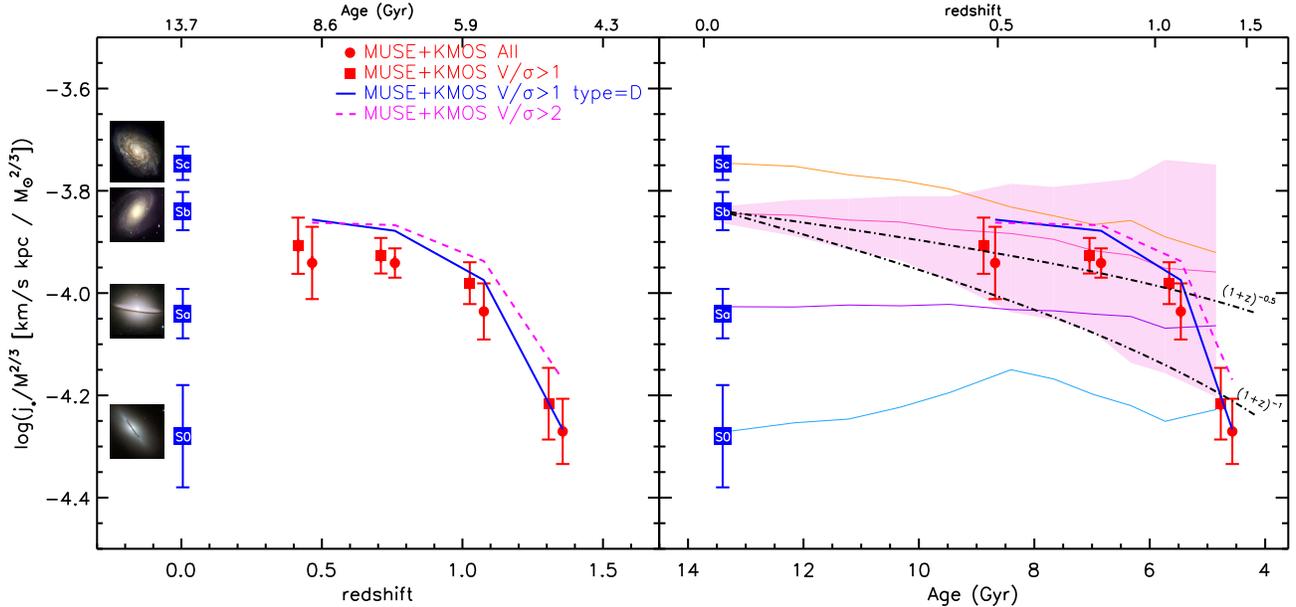,width=7in,angle=0}}
  \caption{{\it Left:} Redshift evolution of the
    $j_\star$\,/\,M$_\star^{2/3}$ from $z$\,=\,0 to $z\sim$\,1.5.  We
    split the $z$\,=\,0 galaxies from \citet{Romanowsky12} in to their
    types (from S0 to Sd).  Data points and their error-bars in all of
    the observational points denote the bootstrap median and scatter
    in the distribution.  {\it Right:} The average evolution of
    individual galaxies in the {\sc eagle} simulation (observational
    data is the same as in the left panel).  We identify all galaxies
    in {\sc eagle} that have angular momentum consistent with late-
    and early-type galaxies at $z$\,=\,0 and use the merger trees to
    measure the ratio of $j_\star$\,/\,M$_\star^{2/3}$ of the main
    progenitor galaxy with redshift.  We show the typical evolution
    and the scatter (68\%) by the shaded region.  In {\sc eagle},
    above $z\sim$\,1.5, the angular momentum of the model early- and
    late-types is similar, but below $z\sim$\,1.5, the ratio of
    $j_\star$\,/\,M$_\star^{2/3}$ grows by $\sim$\,60\% due to a
    combination of gas accretion and outflows (which preferentially
    expel low angular momentum material).  In comparison, the angular
    momentum of the galaxies which end up as early types at $z$\,=\,0
    remains approximately constant over the same period.  We also
    overlay a track of
    $j_\star$\,/\,M$_\star^{2/3}\propto$\,(1\,+\,$z$)$^{-n}$ with
    $n$\,=\,0.5 and $n$\,=\,1.5.  The former represents the evolution
    if the angular momentum grows linearly with time.}
  \label{fig:Jz_z}
\end{figure*}

Since the galaxies in our sample span a range of redshifts, from
$z\sim$\,0.3--1.7, to test how the specific angular momentum evolves
with time, we split our sample in to four redshift bins.  Whilst it
has been instructive to normalise angular momentum by stellar mass
($j_\star$\,=\,$J$\,/\,$M_\star$), the stellar mass is an evolving
quantity, and so we adopt the quantity $j_\star$\,/\,$M_\star^{2/3}$
(or equivalently, $J$\,/\,M$_\star^{5/3}$) and in Fig.~\ref{fig:Jz_z}
we compare the evolution of $j_\star$\,/\,M$_\star^{2/3}$ for our
sample with late- and early-types at $z$\,=\,0.  This figure shows
that there appears to be a trend of increasing specific angular
momentum with decreasing redshift.

%
%

Before interpreting this plot in detail, first we note that Lagos et
al.\ (2016) use {\sc eagle} to show that the redshift evolution of
$j_\star$\,/\,M$_\star^{2/3}$ is sensitive to the mass and star
formation rate limits (see the ``Active'' versus ``Passive''
population in Fig.~12 of Lagos et al.\ 2016).  To test whether our
results are sensitive to selection effects (in particular the evolving
mass limits may result in our observation missing low stellar mass
galaxies at $z\sim$\,1 which are detectable at $z\sim$\,0.3), we
select all of the galaxies from {\sc eagle} between $z$\,=\,0.3--1.5
whose star-formation rates suggest [O{\sc ii}] (or H$\alpha$) emission
line fluxes (calculated using their star-formation rate, redshift and
adopting a typical reddening of A$_{\rm V}$\,=\,0.5) are above $f_{\rm
  lim}$\,=\,1\,$\times$\,10$^{-17}$\,erg\,s$^{-1}$\,cm$^{-2}$.  This
flux limit corresponds approximately to the flux limit of our survey.
We then apply mass cuts of M$_\star$\,=\,0.5, 5, and
20\,$\times$\,10$^{9}$\,M$_\odot$ (which span the lower- median- and
upper-quartiles of the stellar mass range in the observations).  The
stellar mass limits we applied to the {\sc eagle} galaxies (which vary
by a factor 40 from 0.5--20\,$\times$\,10$^9$\,M$_\odot$), result in a
change in the ratio of $j$\,/\,M$_\star^{2/3}$ of (a maximum of)
0.05\,dex.  Thus the trend we see in $j$\,/\,M$^{2/3}$ with redshift
does not appear to be driven by selection biases.

Thus, assuming the majority of the rotationally supported
high-redshift galaxies in our sample continue to evolve towards the
spirals at $z\sim$\,0, Fig.~\ref{fig:Jz_z} suggests a change of
$\Delta (j/M^{2/3})\sim$\,0.4\,dex from $z\sim$\,1 to $z\sim$\,0.
Equivalently, the evolution in $j_\star$\,/\,M$_\star^{2/3}$ is
consistent with $j_\star$\,/\,M$_\star^{2/3}\propto(1+z)^{-n}$ with
$n\sim$\,$-$1.  In the right-hand panel of Fig.~\ref{fig:Jz_z}, we
plot the data in linear time and overlay this redshift evolution.  The
evolution of $j_\star$\,/\,M$_\star^{2/3}\propto(1+z)^{-1}$ is
consistent with that predicted for massive galaxies in {\sc eagle}
(galaxies in halos with masses $M_{\rm
  halo}$\,=\,10$^{11.8-12.3}$\,M$_\odot$; Lagos et al.\ 2016).  In the
figure, we also overlay tracks with
$j_\star$\,/\,M$_\star^{2/3}\propto(1+z)^{-3/2}$ and
$j_\star$\,/\,M$_\star^{2/3}\propto(1+z)^{-1/2}$ to show how the
various predictions compare to the data.

Of course, the assumption that the rotationally supported ``disks'' at
$z\sim$\,1 evolve in to the rotationally supported spirals at
$z\sim$\,0 is difficult to test observationally.  However the model
does allow us to measure how the angular momentum of indivudual
galaxies evolves with time.  To test how the angular momentum of
today's spirals has evolved with time, and in particular what these
evolved from at $z\sim$\,1, we identify all of the galaxies in {\sc
  eagle} whose (final) $j_\star$\,/\,M$_\star^{2/3}$ is consistent
with today's early- and late-types
($j_\star$\,/\,M$^{2/3}_\star$\,=\,$-$3.82\,$\pm$\,0.05 and
$-$4.02\,$\pm$\,0.05 respectively) and trace the evolution of their
angular momentum with redshift (using the main sub-halo progenitor in
each case to trace their dynamics).  We show these evolutionary tracks
in Fig.~\ref{fig:Jz_z}.  In the {\sc eagle} simulation, early-type
galaxies at $z$\,=\,0 have an approximately constant
$j_\star$\,/\,M$_\star^{2/3}$ from $z\sim$\,1.  This is similar to the
findings of Lagos et al.\ (2016) who show that galaxies with
mass-weighted ages $\gsim$\,9\,Gyr have constant $j_\star/M^{2/3}$
with redshift below $z\sim$\,2.  In contrast, the model predicts that
spiral galaxies at $z\sim$\,0 have gradually increased their specific
angular momentum from high redshift, and indeed, for our observed
sample, that the angular momentum of galaxies follows
$j_\star$\,/\,$M_\star^{2/3}\propto(1+z)^{-1}$ (see also
Fig.~\ref{fig:Jz}).  Thus, in the models, the specific angular
momentum of todays spirals was $\sim$\,2.5$\times$ lower that at
$z$\,=\,0.  The increase in $j_\star$\,/\,M$_\star^{2/3}$ has been
attributed to the age at which dark matter halos cease their expansion
(their so-called ``turnaround epoch'') and the fact that star forming
gas at late times has high specific angular momentum which impacts the
disk at large radii (e.g.\ see Fig.~13 of Lagos et al.\ 2016).

It is useful to investigate  the relation between the angular momentum,
stability of the disks and the star formation rate (or star formation
surface density).  As we discussed in \S~\ref{sec:Disk_V_Sigma}, the
stability of a gas disk against clump formation is quantified by the
Toomre parameter, $Q$.  Recently, \citet{Obreschkow15} suggested that
the low angular momentum of high-redshift galaxies is the dominant
driver of the formation of ``clumps'', which hence leads to
clumpy/disturbed mophologies and intense star formation.  As the
specific angular momentum increases with decreasing redshift, the
disk-average average Toomre $Q$ becomes greater than unity and the
disk becomes globally stable.

To test whether this is consistent with
the galaxies in our sample, we select all the rotationally-supported
galaxies from our MUSE and KMOS survey that have stellar masses
greater than M$_\star$\,=\,10$^{10}$\,M$_{\odot}$, and split the
sample in to galaxies above and below
$j_\star$\,/\,M$_\star^{2/3}$\,=\,10$^{2.5}$\,km\,s$^{-1}$\,kpc\,M$_\odot^{-2/3}$
(we only consider galaxies above M$_\star$\,=\,10$^{10}$\,M$_{\odot}$
since these are well resolved in our data).  For these two
sub-samples, we derive $Q$\,=\,1.10\,$\pm$\,0.18 for the galaxies with
the highest $j_\star$\,/\,M$_\star^{2/3}$ and $Q$\,=\,0.53\,$\pm$\,0.22 for those
galaxies with the lowest $j_\star$\,/\,M$_\star^{2/3}$.  This is not a particularly
surprising result since the angular momentum and Toomre $Q$ are both a
strong function of rotational velocity and radius However, it is
interesting to note that the average star formation rate and star
formation surface densities of these two subsets of high and low
$j_\star$\,/\,M$^{2/3}$ are also markedly different.
For the galaxies
above the $j_\star$\,/\,M$^{2/3}$ sequence at this mass, the star
formation rates and star-formation surface densities are
SFR\,=\,8\,$\pm$\,4\,M$_\odot$\,yr$^{-1}$ and $\Sigma_{\rm
  SFR}$\,=\,123\,$\pm$\,23\,M$_\odot$\,yr$^{-1}$\,kpc$^2$
respectively.  In comparison, the galaxies below the sequence have
higher rates, with SFR\,=\,21\,$\pm$\,4\,M$_\odot$\,yr$^{-1}$ and
$\Sigma_{\rm SFR}$\,=\,206\,$\pm$\,45\,M$_\odot$\,yr$^{-1}$\,kpc$^2$
respectively.  In this comparison, the star-formation rates are the
most illustrative indication of the difference in sub-sample
properties since they are independent of $j_\star$, stellar mass and
size.

\begin{figure*}
  \centerline{\psfig{file=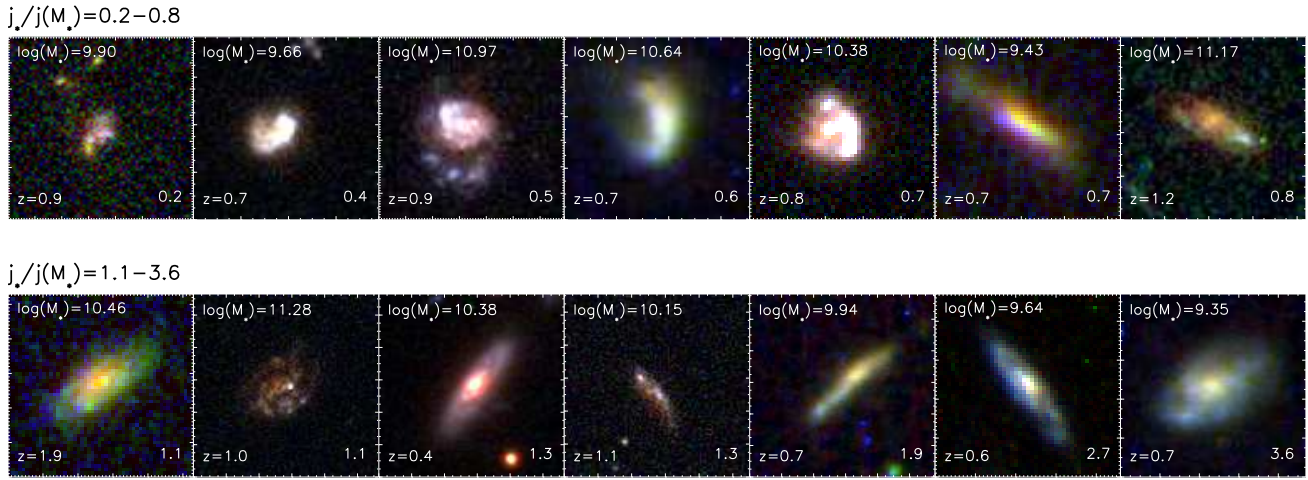,width=7in,angle=90}}
  \caption{\emph{HST} colour images of fourteen galaxies from
    Fig.~\ref{fig:Jz} whose specific angular momentum ($j_\star$) are
    below the $j_\star$--M$_\star$ sequence (upper row) and below the
    $j_\star$--M$_\star$ sequence (lower row).  The galaxies shown in
    this figure all have stellar masses
    $>$\,2\,$\times$\,10$^9$\,M$_\odot$, with similar stellar masses
    and redshift distributions
    (log$_{10}$(M$_\star$\,/\,M$_\star$\,=\,10.3\,$\pm$\,0.4 and
    10.2\,$\pm$\,0.3 and $z$\,=\,0.78\,$\pm$\,0.10 and
    0.74\,$\pm$\,0.11 respectively for the upper and lower rows).  The
    stellar masses and redshifts are given in the upper and lower
    left-hand corners respectively.  The value in the lower-right hand
    corner of each image is the fractional offset from the
    $j_\star$--M$_\star$ sequence in Fig.~\ref{fig:Jz} (i.e.\ a value
    of 0.2 means that galaxy has a specific angular momentum that is
    5$\times$ lower than the $j_\star$--$M_\star$ sequence given its
    stellar mass).  These images demonstrate that galaxies with lower
    specific angular momentum (at fixed mass) are those with more
    disturbed morphologies and larger and brighter clumps (upper row),
    whilst the galaxies with high angular momentum are those with
    morphologies that more cloesly resemble spiral galaxies (lower
    row).  The with low specific angular momentum are also dynamically
    unstable, with Toomre $Q$\,=\,0.53\,$\pm$\,0.22 compared to those
    with high specific angular momentum which have Toomre
    $Q$\,=\,1.10\,$\pm$\,0.18.  Together, this demonstrates that the
    disk stability and morphology of the galaxies is strongly
    correlated with the angular momentum of the gas disk.}
  \label{fig:J_morph}
\end{figure*}

Since a large fraction of our sample have been observed usign
\emph{HST}, we can also investigate the morphologies of those galaxies
above and below the specific angular momentum--stellar mass sequence.
In Fig.~\ref{fig:J_morph} we show \emph{HST} colour images of fourteen
galaxies, seven each with specific angular momentum ($j_\star$) that
are above or below the $j_\star$--M$_\star$ sequence.  We select
galaxies for this plot which are matched in redshift and stellar mass
(all have stellar masses $>$\,2\,$\times$\,10$^9$\,M$_\odot$, with
medians of log$_{10}$(M$_\star$\,/\,M$_\star$)\,=\,10.3\,$\pm$\,0.4
and 10.2\,$\pm$\,0.3 and $z$\,=\,0.78\,$\pm$\,0.10 and
0.74\,$\pm$\,0.11 respectively for the upper and lower rows).  Whilst
a full morphological analysis is beyond the scope of this paper, it
appears from this plot that the galaxies with higher specific angular
momentum (at fixed mass) are those with more established (smoother)
disks.  In contrast, the galaxies with lower angular momentum are
those with morphologies that are either more compact, more disturbed
morphologies and/or larger and brighter clumps.

%
%

Taken together, these results suggest that at $z\sim$\,1, galaxies follow a
similar scaling between mass and specific angular momentum as those at
$z\sim$\,0.  However, at high masses ($>$M$_\star$ at $z\sim$\,1)
star-forming galaxies have lower specific angular momentum (by a
factor $\sim$\,2.5) than a mass matched sample at $z\sim$\,0, and we
do not find any high-redshift galaxies with specific angular momentum
as high as those in local spirals.  From their Toomre stability and
star formation surface densities, the most unstable disks have the
lowest specific angular momentum, asymmetric morphologies and
highest star formation rate surface densities \citep[see
  also][]{Obreschkow15}.  Galaxies with higher specific angular
momentum appear to be more stable, with smoother (disk-like)
morphologies. 

\begin{figure}
  \centerline{\psfig{file=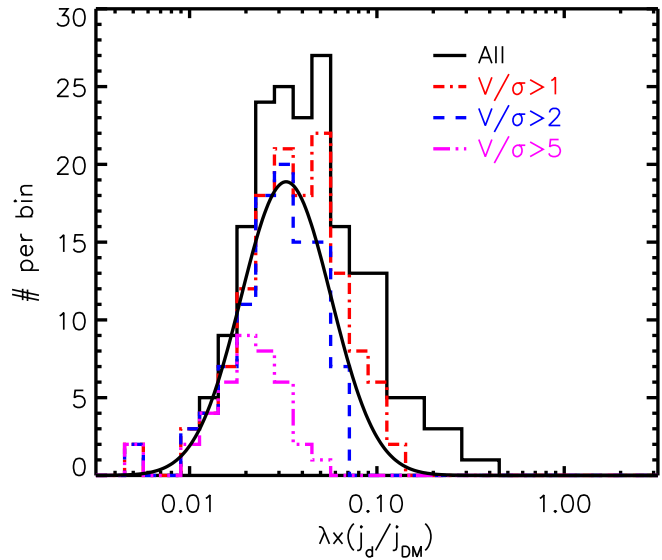,width=3.5in,angle=0}}
  \caption{The distribution of spin ($\lambda\times$\,($j_{\rm
      d}$\,/\,$j_{\rm DM}$)) for the galaxies in our sample.  We split
    the sample in to four catagories: All rotationally supported
    systems, and those with V\,/\,$\sigma>$\,1, 2 and 5.  We fit the
    distribution of galaxies with V\,/\,$\sigma>$\,2 with a power-law
    relation to derive best-fit parameters
    $\lambda'$\,=\,0.040\,$\pm$\,0.002 and
    $\sigma$\,=\,0.045\,$\pm$\,0.005.}
\label{fig:spin}
\end{figure}

Finally, we calculate the distribution of baryonic spins for our
sample.  The spin typically refers to the fraction of centrifugal
support for the halo.  Both linear theory and N-body simulations have
suggested that halos have spins that follow approximately log-normal
distributions with average value $\lambda_{\rm DM}$\,=\,0.035
\citep{Bett07} (i.e.\ only $\sim$3.5\% of the dynamical support of a
halo is centrifugal, the rest comes from dispersion).  To estimate how
the disk and halo angular momentum are related, we calculate the spin
of the disk, $\lambda$ as $\lambda$\,=\,$\sqrt{2}$\,/\,0.1\,$R_{\rm
  d}$\,$H(z)$\,/\,V(3\,$R_{\rm d}$) where $H(z)$\,=\,$H_{\rm
  0}(\Omega_{\rm \Lambda,0}+\Omega_{\rm m,0}(1+z)^{3})^{0.5}$.  This
is the simplest approach that assumes the galaxy is embedded inside an
isothermal spherical cold-dark matter halos
\citep[e.g.\ ][]{White84,Mo98} which are truncated at the virial
radius (\citealt{Peebles69}; see \citealt{Burkert16} for a discussion
for the results from adopting more complex halo profiles).  In
Fig.~\ref{fig:spin} we plot the distribution of
$\lambda\times$\,($j_{\rm d}$\,/\,$j_{\rm DM}$) for our sample.  If
the initial halo and baryonic angular momentum are similar,
i.e. $j_{\rm DM}\simeq j_{d}$, this quantity reflects the fraction of
angular momentum lost during the formation of $z\sim$\,1 star-forming
galaxies.  In this figure, we split the sample in to four catagories:
all galaxies with disk-like dynamics with V\,/\,$\sigma>$\,1, 2 and 5.
We fit these distribution with a log-normal power-law distribution,
deriving best-fit parameters in [$\lambda',\sigma$] of
[0.040\,$\pm$\,0.002, 0.45\,$\pm$\,0.05], [0.041\,$\pm$\,0.002,
  0.42\,$\pm$\,0.05] and [0.068\,$\pm$\,0.002, 0.50\,$\pm$\,0.04]
respectively.

%
%

An alternative approach (see also \citealt{Harrison17}) is to assume
the spin for the baryons of $\lambda'$\,=\,0.035 and calculate the
fraction of angular momentum that has been retained (assuming $j_{\rm
  DM}\simeq j_{d}$ initially).  For the galaxies that appear to be
rotationally supported with ratios of $V$\,/\,$\sigma>$1, 2 and 5 we
derive median values of $j_{\rm d}$\,/\,$j_{\rm
  DM}\sim$\,1.18\,$\pm$\,0.10, 0.95\,$\pm$\,0.06 and 0.70\,$\pm$\,0.05
(bootstrap errors).  Since these spins are similar to the halo
($\lambda$\,=\,0.035) this suggests that the angular momentum of
``rotationally supported'' galaxies at $z\sim$\,1 broadly follows that
expected from theoretical expectation from the halo, with most of the
angular momentum retained during the (initial) collapse.
Equivalently, for the galaxies with the highest ratio of
$V$\,/\,$\sigma$ (which are also those with the highest specific
angular momentum and latest morphological types; see
Fig.~\ref{fig:J_morph}), the fraction of angular momentum retained
must be $\gsim$\,70\%.

\section{Conclusions}
\label{sec:conclusions}

Exploiting MUSE and KMOS observations, we study the dynamics of 405
star-forming galaxies across the redshift range $z$\,=\,0.28--1.65,
with a median redshift of $z$\,=\,0.84.  From estimates of their
stellar masses and star formation rates, our sample appear to be
representative of the star-forming ``main-sequence'' from
$z$\,=\,0.3--1.7, with ranges of
SFR\,=\,0.1--30\,M$_{\odot}$\,yr$^{-1}$ and
M$_{\star}$\,=\,10$^8$--10$^{11}$\,M$_{\odot}$.

\noindent Our main results are summarised as:\\

$\bullet$ From the dynamics and morphologies of the galaxies in the
sample, 49\,$\pm$\,4\% appear to be rotationally supported;
24\,$\pm$\,3\% are unresolved; and only $\sim$\,5\,$\pm$\,2\%
appear to be major mergers.  The remainder appear to be irregular (or
perhaps face-on) systems.  Our estimate of the ``disk'' fraction in
this sample is consistent with other dynamical studies over a similar
redshift range which have also found that rotationally supported
systems make up $\sim$\,40--70\% of the star-forming population.

$\bullet$ We measure half light sizes of the galaxies in both the
broad-band continuum images (using \emph{HST} imaging in many cases)
and in the nebular emission lines.  The nebular emission line sizes
are typically a factor of 1.18\,$\pm$\,0.03\,$\times$ larger than the
continuum sizes.  This is consistent with recent results from the
3-D\emph{HST} survey which has also shown that the nebular emission
from $\sim$\,$L^{\star}$ star-forming galaxies at $z\sim$\,1 are
systematically more extended than the stellar continuum. 

$\bullet$ For those galaxies whose dynamics resemble rotationally
supported systems, we simultaneously fit the imaging and dynamics with
a disk\,+\,halo model to derive the best-fit structural parameters
(such as disk inclination, position angle, [$x$\,/\,$y$] center, disk
mass, disk size, dark matter core radius and density).  The dynamical
and morphological major axes are typically misaligned by
$\Delta$\,PA\,=\,9.5\,$\pm$\,0.5$^{\circ}$, which we attribute to the
dynamical ``settling'' of the gas and stars as the disks evolve.



$\bullet$ We combine the inclination-corrected rotational velocities
with the galaxy sizes and intrinsic velocity dispersions to
investigate the global stability of the gas disks.  For the galaxies
that are classified as rotationally supported, we derive a median
Toomre $Q$ of $Q$\,=\,0.80\,$\pm$\,0.10.  This is consistent with
numerical models that predict that in high-redshift, gas rich galaxies
the disks are maintained at the marginally stable threshold due to the
feedback from stellar winds which arrest collapse
\citep[e.g.\ ][]{Hopkins12b}.

$\bullet$ We use the galaxy sizes, rotation speeds and stellar masses
to investigate how the specific angular momentum of gas disks evolves
with cosmic time.  We show that the galaxies in our sample (which have
a median redshift of $z$\,=\,0.84\,$\pm$\,0.03) follow a similar
scaling between stellar mass and specific angular momentum as local
galaxies.  Fitting the data over the stellar mass range
M$_\star$\,=\,10$^{8.5}$--10$^{11.5}$\,M$_\odot$ suggests
$j_\star\propto$\,M$_\star^{\rm q}$ with $q$\,=\,0.6\,$\pm$\,0.1.
However, at $z$\,=\,1, we do not find any galaxies with specific
angular momentum as high as those of local spirals.  Thus, the most
massive star-forming disks at $z\sim$\,0 must have increased their
specific angular momentum (by a factor $\sim$\,3) between $z\sim$\,1
and $z\sim$\,0.

$\bullet$ To account for the evolving stellar masses of galaxies, we
measure the ratio of $j_\star$\,/\,M$_\star^{2/3}$ and split our
observed sample in to four redshift bins between $z$\,=\,0.3 and
$z$\,=\,1.5.  For a star-formation selected (and mass limited) sample,
we show that the specific angular momentum evolves with redshift as
$j_\star\propto$\,M$_\star^{2/3}(1+z)^{-1}$, which is similar to that
predicted by the latest numerical models, which also suggest that
spiral galaxies at $z\sim$\,0 appear to have gradually accreted their
specific angular momentum from high redshift (in contrast to
``passive'' galaxies at $z\sim$\,0 which, on average, have near
constant specific angular momentum between $z\sim$\,0 and $z\sim$\,1).

$\bullet$ Combining the measurements of the angular momentum, star
formation surface density and disk stability, we show that galaxies
with stellar masses greater than M$_\star$\,=\,10$^{10}$\,M$_{\odot}$
with the highest $j_\star$\,/\,M$_\star^{2/3}$ are the most stable,
with disks with Toomre $Q$\,=\,1.20\,$\pm$\,0.20, compared to
$Q$\,=\,0.51\,$\pm$\,0.17 for galaxies above and below
$j_\star$\,/\,M$_\star^{2/3}$\,=\,10$^{2.5}$\,km\,s$^{-1}$\,kpc\,M$_\odot^{-2/3}$
respectively.  Since $j_\star$ and $Q$ are both functions of size and
rotational velocity, we also measure the average star formation rate
and star formation surface densities of these two subsets of high and
low $j_\star$\,/\,M$^{2/3}$ galaxies.  These sub-samples are markedly
different, with a median $\Sigma_{\rm
  SFR}$\,=\,123\,$\pm$\,23\,M$_\odot$\,yr$^{-1}$ and $\Sigma_{\rm
  SFR}$\,=\,206\,$\pm$\,45\,M$_\odot$\,yr$^{-1}$ for those galaxies
above and below the fiducial $j_\star$\,/\,M$_\star^{2/3}$ relation
respectively.  In terms of star formation rates alone, there is a
similar difference, with SFR\,=\,8\,$\pm$\,4\,M$_\odot$\,yr$^{-1}$ and
SFR\,=\,21\,$\pm$\,4\,M$_\odot$\,yr$^{-1}$ above and below the
sequence respectively.

$\bullet$ At a fixed mass, we show that galaxies with high specific
angular momentum ($j_\star$) (i.e.\ those above the
$j_\star$--M$_\star$ ) relation are those with morphologies that more
closely resemble spiral galaxies, with bigger bulges and smoother
disks.  In contrast, galaxies with lower specific angular momentum (at
fixed mass) are those with more disturbed, asymmetric morphologies,
larger and brighter clumps.

$\bullet$ Finally, we show that the distribution of spins for the
rotationally supported galaxies in our sample is similar to that
expected for the halos.  For exampe, for galaxies that have disk like
dynamics and $V$\,/\,$\sigma>$\,2 we derive
$\lambda'$\,=\,0.040\,$\pm$\,0.002 and $\sigma$\,=\,0.45\,$\pm$\,0.05.
This suggests that the angular momentum of ``rotationally supported''
galaxies at $z\sim$\,1 broadly follows that expected from theoretical
expectation from the halo, with most of the angular momentum retained
during the (initial) collapse.

Overall, our results show that star forming disks at $z\sim$\,1 have
lower specific angular momentum than a stellar mass matched sample at
$z\sim$\,0.  At high redshift, the fraction of rotationally supported
``disk'' galaxies is high, yet most of these galaxies appear
irregular/clumpy.  This appears to be due to the high low angular
momentum which results in globally unstable, turbulent systems.
Indeed, specific angular momentum appears to play a major role in
defining the disk stability, star formation surface densities and
morphology.  As the specific angular momentum of growing disks
increases below $z\sim$\,1, the galaxy disks must evolve from globally
unstable clumpy, turbulent systems in to stable, flat regular spirals.

\section*{Acknowledgments}

We thank the anonymous referee for their constructive report which
significantly improved the content and clarity of this work.  AMS
gratefully acknowledges an STFC Advanced Fellowship through grant
number ST/H005234/1, the Leverhume foundation and STFC grant
ST/L00075X/1.  IRS acknowledges support from STFC (ST/L00075X/1), the
ERC Advanced Investigator programme 321334-DUSTYGAL and a Royal
Society/Wolfson Merit Award.  JS and RB acknowledge ERC advanced
grants 278594-GasAroundGalaxies and 339659-MUSICOS respectively.  DS
acknowledges financial support from the Netherlands organisation for
scientific reserach (NWO) through a Vani fellowship and from a FCT
investigator starting grant (IF/01154/2012/CP0189/CT0010).  RAC
acknowledges a Royal Society University Research Fellowship.

\bibliographystyle{apj}

\begin{thebibliography}{90}
\expandafter\ifx\csname natexlab\endcsname\relax\def\natexlab#1{#1}\fi

\bibitem[{{Bacon} {et~al.}(2010){Bacon}, {Accardo}, {Adjali}, {Anwand},
  {Bauer}, {Biswas}, {Blaizot}, \& et~al.}]{Bacon10SPIE}
{Bacon}, R., {Accardo}, M., {Adjali}, L., {Anwand}, H., {Bauer}, S., {Biswas},
  I., {Blaizot}, J., \& et~al., B. 2010, in Proceedings of SPIE, Vol. 7735,
  Ground-based and Airborne Instrumentation for Astronomy III, 773508

\bibitem[{{Bacon} {et~al.}(2015){Bacon}, {Brinchmann}, {Richard}, {Contini},
  {Drake}, {Franx}, {Tacchella}, \& {Vernet, J. et al.}}]{Bacon15}
{Bacon}, R., {Brinchmann}, J., {Richard}, J., {Contini}, T., {Drake}, A.,
  {Franx}, M., {Tacchella}, S., \& {Vernet, J. et al.} 2015, \aap, 575, A75

\bibitem[{{Barnes} \& {Efstathiou}(1987)}]{Barnes87}
{Barnes}, J. \& {Efstathiou}, G. 1987, \apj, 319, 575

\bibitem[{{Bell} {et~al.}(2004){Bell}, {McIntosh}, {Barden}, {Wolf},
  {Caldwell}, {Rix}, {Beckwith}, \& {Borch, A. et al.}}]{Bell04}
{Bell}, E.~F., {McIntosh}, D.~H., {Barden}, M., {Wolf}, C., {Caldwell},
  J.~A.~R., {Rix}, H.-W., {Beckwith}, S.~V.~W., \& {Borch, A. et al.} 2004,
  \apjl, 600, L11

\bibitem[{{Bertin} \& {Arnouts}(1996)}]{Bertin96}
{Bertin}, E. \& {Arnouts}, S. 1996, \aaps, 117, 393

\bibitem[{{Bertola} \& {Capaccioli}(1975)}]{Bertola75}
{Bertola}, F. \& {Capaccioli}, M. 1975, \apj, 200, 439

\bibitem[{{Bett} {et~al.}(2007){Bett}, {Eke}, {Frenk}, {Jenkins}, {Helly}, \&
  {Navarro}}]{Bett07}
{Bett}, P., {Eke}, V., {Frenk}, C.~S., {Jenkins}, A., {Helly}, J., \&
  {Navarro}, J. 2007, \mnras, 376, 215

\bibitem[{{Bolzonella} {et~al.}(2000){Bolzonella}, {Miralles}, \& {Pell{\'
  o}}}]{Bolzonella00}
{Bolzonella}, M., {Miralles}, J.-M., \& {Pell{\' o}}, R. 2000, \aap, 363, 476

\bibitem[{{Brammer} {et~al.}(2008){Brammer}, {van Dokkum}, \&
  {Coppi}}]{Brammer08EASY}
{Brammer}, G.~B., {van Dokkum}, P.~G., \& {Coppi}, P. 2008, \apj, 686, 1503

\bibitem[{{Bruce} {et~al.}(2014){Bruce}, {Dunlop}, {McLure}, {Cirasuolo},
  {Buitrago}, {Bowler}, {Targett}, \& {Bell, E.~F. et al.}}]{Bruce14}
{Bruce}, V.~A., {Dunlop}, J.~S., {McLure}, R.~J., {Cirasuolo}, M., {Buitrago},
  F., {Bowler}, R.~A.~A., {Targett}, T.~A., \& {Bell, E.~F. et al.} 2014,
  \mnras, 444, 1660

\bibitem[{{Buitrago} {et~al.}(2013){Buitrago}, {Trujillo}, {Conselice}, \&
  {H{\"a}u{\ss}ler}}]{Buitrago13}
{Buitrago}, F., {Trujillo}, I., {Conselice}, C.~J., \& {H{\"a}u{\ss}ler}, B.
  2013, \mnras, 428, 1460

\bibitem[{{Bundy} {et~al.}(2005){Bundy}, {Ellis}, \& {Conselice}}]{Bundy05}
{Bundy}, K., {Ellis}, R.~S., \& {Conselice}, C.~J. 2005, \apj, 625, 621

\bibitem[{{Burkert}(1995)}]{Burkert95}
{Burkert}, A. 1995, \apjl, 447, L25

\bibitem[{{Burkert}(2009)}]{Burkert09}
{Burkert}, A. 2009, in Astronomical Society of the Pacific Conference Series,
  Vol. 419, Galaxy Evolution: Emerging Insights and Future Challenges, ed.
  S.~{Jogee}, I.~{Marinova}, L.~{Hao}, \& G.~A. {Blanc}, 3

\bibitem[{{Burkert} {et~al.}(2015){Burkert}, {F{\"o}rster Schreiber}, {Genzel},
  {Lang}, {Tacconi}, {Wisnioski}, {Wuyts}, \& {Bandara, K. et al.}}]{Burkert16}
{Burkert}, A., {F{\"o}rster Schreiber}, N.~M., {Genzel}, R., {Lang}, P.,
  {Tacconi}, L.~J., {Wisnioski}, E., {Wuyts}, S., \& {Bandara, K. et al.} 2015,
  ArXiv

\bibitem[{{Calzetti} {et~al.}(2000){Calzetti}, {Armus}, {Bohlin}, {Kinney},
  {Koornneef}, \& {Storchi-Bergmann}}]{Calzetti00}
{Calzetti}, D., {Armus}, L., {Bohlin}, R.~C., {Kinney}, A.~L., {Koornneef}, J.,
  \& {Storchi-Bergmann}, T. 2000, \apj, 533, 682

\bibitem[{{Catelan} \& {Theuns}(1996)}]{Catelan96}
{Catelan}, P. \& {Theuns}, T. 1996, \mnras, 282, 436

\bibitem[{{Ceverino} {et~al.}(2010){Ceverino}, {Dekel}, \&
  {Bournaud}}]{Ceverino10}
{Ceverino}, D., {Dekel}, A., \& {Bournaud}, F. 2010, \mnras, 404, 2151

\bibitem[{{Ciardullo} {et~al.}(2013){Ciardullo}, {Gronwall}, {Adams}, {Blanc},
  {Gebhardt}, {Finkelstein}, {Jogee}, {Hill}, {Drory}, {Hopp}, {Schneider},
  {Zeimann}, \& {Dalton}}]{Ciadrdullo13}
{Ciardullo}, R., {Gronwall}, C., {Adams}, J.~J., {Blanc}, G.~A., {Gebhardt},
  K., {Finkelstein}, S.~L., {Jogee}, S., {Hill}, G.~J., {Drory}, N., {Hopp},
  U., {Schneider}, D.~P., {Zeimann}, G.~R., \& {Dalton}, G.~B. 2013, \apj, 769,
  83

\bibitem[{{Cole} \& {Lacey}(1996)}]{Cole96}
{Cole}, S. \& {Lacey}, C. 1996, \mnras, 281, 716

\bibitem[{{Conselice} {et~al.}(2011){Conselice}, {Bluck}, {Ravindranath},
  {Mortlock}, {Koekemoer}, {Buitrago}, {Gr{\"u}tzbauch}, \&
  {Penny}}]{Conselice11}
{Conselice}, C.~J., {Bluck}, A.~F.~L., {Ravindranath}, S., {Mortlock}, A.,
  {Koekemoer}, A.~M., {Buitrago}, F., {Gr{\"u}tzbauch}, R., \& {Penny}, S.~J.
  2011, \mnras, 417, 2770

\bibitem[{{Contini} {et~al.}(2015){Contini}, {Epinat}, {Bouch{\'e}},
  {Brinchmann}, {Boogaard}, {Ventou}, {Bacon}, \& {Richard, J. et
  al.}}]{Contini16}
{Contini}, T., {Epinat}, B., {Bouch{\'e}}, N., {Brinchmann}, J., {Boogaard},
  L.~A., {Ventou}, E., {Bacon}, R., \& {Richard, J. et al.} 2015, ArXiv
  e-prints

\bibitem[{{Courteau}(1997)}]{Courteau97}
{Courteau}, S. 1997, \aj, 114, 2402

\bibitem[{{Crain} {et~al.}(2015){Crain}, {Schaye}, {Bower}, {Furlong},
  {Schaller}, {Theuns}, {Dalla Vecchia}, \& {Frenk, C.~S.}}]{Crain15}
{Crain}, R.~A., {Schaye}, J., {Bower}, R.~G., {Furlong}, M., {Schaller}, M.,
  {Theuns}, T., {Dalla Vecchia}, C., \& {Frenk, C.~S.} 2015, ArXiv e-prints

\bibitem[{{Danovich} {et~al.}(2015){Danovich}, {Dekel}, {Hahn}, {Ceverino}, \&
  {Primack}}]{Danovich15}
{Danovich}, M., {Dekel}, A., {Hahn}, O., {Ceverino}, D., \& {Primack}, J. 2015,
  \mnras, 449, 2087

\bibitem[{{Dekel} {et~al.}(2009){Dekel}, {Sari}, \& {Ceverino}}]{Dekel09b}
{Dekel}, A., {Sari}, R., \& {Ceverino}, D. 2009, \apj, 703, 785

\bibitem[{{Emsellem} {et~al.}(2007){Emsellem}, {Cappellari}, {Krajnovi{\'c}},
  {van de Ven}, {Bacon}, {Bureau}, {Davies}, \& {de Zeeuw, P, et
  al.}}]{Emsellem07}
{Emsellem}, E., {Cappellari}, M., {Krajnovi{\'c}}, D., {van de Ven}, G.,
  {Bacon}, R., {Bureau}, M., {Davies}, R.~L., \& {de Zeeuw, P, et al.} 2007,
  \mnras, 379, 401

\bibitem[{{Epinat} {et~al.}(2012){Epinat}, {Tasca}, {Amram}, {Contini}, {Le
  F{\`e}vre}, {Queyrel}, {Vergani}, {Garilli}, {Kissler-Patig}, {Moultaka},
  {Paioro}, {Tresse}, {Bournaud}, {L{\'o}pez-Sanjuan}, \& {Perret}}]{Epinat12}
{Epinat}, B., {Tasca}, L., {Amram}, P., {Contini}, T., {Le F{\`e}vre}, O.,
  {Queyrel}, J., {Vergani}, D., {Garilli}, B., {Kissler-Patig}, M., {Moultaka},
  J., {Paioro}, L., {Tresse}, L., {Bournaud}, F., {L{\'o}pez-Sanjuan}, C., \&
  {Perret}, V. 2012, \aap, 539, A92

\bibitem[{{Fall}(1983)}]{Fall83}
{Fall}, S.~M. 1983, in IAU Symposium, Vol. 100, Internal Kinematics and
  Dynamics of Galaxies, ed. E.~{Athanassoula}, 391--398

\bibitem[{{Fall} \& {Romanowsky}(2013)}]{Fall13}
{Fall}, S.~M. \& {Romanowsky}, A.~J. 2013, \apjl, 769, L26

\bibitem[{{Ferguson} {et~al.}(2004){Ferguson}, {Dickinson}, {Giavalisco},
  {Kretchmer}, {Ravindranath}, {Idzi}, {Taylor}, {Conselice}, {Fall},
  {Gardner}, {Livio}, {Madau}, {Moustakas}, {Papovich}, {Somerville},
  {Spinrad}, \& {Stern}}]{Ferguson04}
{Ferguson}, H.~C., {Dickinson}, M., {Giavalisco}, M., {Kretchmer}, C.,
  {Ravindranath}, S., {Idzi}, R., {Taylor}, E., {Conselice}, C.~J., {Fall},
  S.~M., {Gardner}, J.~P., {Livio}, M., {Madau}, P., {Moustakas}, L.~A.,
  {Papovich}, C.~M., {Somerville}, R.~S., {Spinrad}, H., \& {Stern}, D. 2004,
  \apjl, 600, L107

\bibitem[{{F{\"o}rster Schreiber} {et~al.}(2009){F{\"o}rster Schreiber},
  {Genzel}, {Bouch{\'e}}, {Cresci}, {Davies}, {Buschkamp}, {Shapiro}, \&
  {Tacconi, L.~J.. et al.}}]{ForsterSchreiber09}
{F{\"o}rster Schreiber}, N.~M., {Genzel}, R., {Bouch{\'e}}, N., {Cresci}, G.,
  {Davies}, R., {Buschkamp}, P., {Shapiro}, K., \& {Tacconi, L.~J.. et al.}
  2009, \apj, 706, 1364

\bibitem[{{Franx} {et~al.}(1991){Franx}, {Illingworth}, \& {de
  Zeeuw}}]{Franx91}
{Franx}, M., {Illingworth}, G., \& {de Zeeuw}, T. 1991, \apj, 383, 112

\bibitem[{{Freeman}(1970)}]{Freeman70}
{Freeman}, K.~C. 1970, \apj, 160, 811

\bibitem[{{Furlong} {et~al.}(2015){Furlong}, {Bower}, {Theuns}, {Schaye},
  {Crain}, {Schaller}, {Dalla Vecchia}, {Frenk}, {McCarthy}, {Helly},
  {Jenkins}, \& {Rosas-Guevara}}]{Furlong15}
{Furlong}, M., {Bower}, R.~G., {Theuns}, T., {Schaye}, J., {Crain}, R.~A.,
  {Schaller}, M., {Dalla Vecchia}, C., {Frenk}, C.~S., {McCarthy}, I.~G.,
  {Helly}, J., {Jenkins}, A., \& {Rosas-Guevara}, Y.~M. 2015, \mnras, 450, 4486

\bibitem[{{Gallagher} \& {Hunter}(1984)}]{Gallagher84}
{Gallagher}, III, J.~S. \& {Hunter}, D.~A. 1984, \araa, 22, 37

\bibitem[{{Geach} {et~al.}(2008){Geach}, {Smail}, {Best}, {Kurk}, {Casali},
  {Ivison}, \& {Coppin}}]{Geach08}
{Geach}, J.~E., {Smail}, I., {Best}, P.~N., {Kurk}, J., {Casali}, M., {Ivison},
  R.~J., \& {Coppin}, K. 2008, \mnras, 388, 1473

\bibitem[{{Genel} {et~al.}(2015){Genel}, {Fall}, {Hernquist}, {Vogelsberger},
  {Snyder}, {Rodriguez-Gomez}, {Sijacki}, \& {Springel}}]{Genel15}
{Genel}, S., {Fall}, S.~M., {Hernquist}, L., {Vogelsberger}, M., {Snyder},
  G.~F., {Rodriguez-Gomez}, V., {Sijacki}, D., \& {Springel}, V. 2015, \apjl,
  804, L40

\bibitem[{{Genzel} {et~al.}(2011){Genzel}, {Newman}, {Jones}, {F{\"o}rster
  Schreiber}, {Shapiro}, {Genel}, {Lilly}, \& {Renzini, A. et al.}}]{Genzel11}
{Genzel}, R., {Newman}, S., {Jones}, T., {F{\"o}rster Schreiber}, N.~M.,
  {Shapiro}, K., {Genel}, S., {Lilly}, S.~J., \& {Renzini, A. et al.} 2011,
  \apj, 733, 101

\bibitem[{{Giavalisco} {et~al.}(1996){Giavalisco}, {Steidel}, \&
  {Macchetto}}]{Giavalisco96}
{Giavalisco}, M., {Steidel}, C.~C., \& {Macchetto}, F.~D. 1996, \apj, 470, 189

\bibitem[{{Gnerucci} {et~al.}(2011){Gnerucci}, {Marconi}, {Cresci}, {Maiolino},
  {Mannucci}, {Calura}, {Cimatti}, \& {Cocchia, F. et al.}}]{Gnerucci11}
{Gnerucci}, A., {Marconi}, A., {Cresci}, G., {Maiolino}, R., {Mannucci}, F.,
  {Calura}, F., {Cimatti}, A., \& {Cocchia, F. et al.} 2011, \aap, 528, A88

\bibitem[{{Guo} {et~al.}(2009){Guo}, {McIntosh}, {Mo}, {Katz}, {van den Bosch},
  {Weinberg}, {Weinmann}, {Pasquali}, \& {Yang}}]{Guo09}
{Guo}, Y., {McIntosh}, D.~H., {Mo}, H.~J., {Katz}, N., {van den Bosch}, F.~C.,
  {Weinberg}, M., {Weinmann}, S.~M., {Pasquali}, A., \& {Yang}, X. 2009,
  \mnras, 398, 1129

\bibitem[{{Harrison} {et~al.}(2017){Harrison}, {Johnson}, {Swinbank}, {Stott},
  {Bower}, {Smail}, {Tiley}, \& {Bunker, A.~J. et al.}}]{Harrison17}
{Harrison}, C.~M., {Johnson}, H.~L., {Swinbank}, A.~M., {Stott}, J.~P.,
  {Bower}, R.~G., {Smail}, I., {Tiley}, A.~L., \& {Bunker, A.~J. et al.} 2017,
  ArXiv e-prints

\bibitem[{{Hopkins}(2012)}]{Hopkins12b}
{Hopkins}, P.~F. 2012, \mnras, 423, 2016

\bibitem[{{Husband} {et~al.}(2015){Husband}, {Bremer}, {Stanway}, \&
  {Lehnert}}]{Husband15}
{Husband}, K., {Bremer}, M.~N., {Stanway}, E.~R., \& {Lehnert}, M.~D. 2015,
  \mnras, 452, 2388

\bibitem[{{Kennicutt}(1998)}]{KS98}
{Kennicutt}, Jr., R.~C. 1998, \apj, 498, 541

\bibitem[{{Khostovan} {et~al.}(2015){Khostovan}, {Sobral}, {Mobasher}, {Best},
  {Smail}, {Stott}, {Hemmati}, \& {Nayyeri}}]{Khostovan15}
{Khostovan}, A.~A., {Sobral}, D., {Mobasher}, B., {Best}, P.~N., {Smail}, I.,
  {Stott}, J.~P., {Hemmati}, S., \& {Nayyeri}, H. 2015, ArXiv e-prints

\bibitem[{{Kriek} {et~al.}(2009){Kriek}, {van Dokkum}, {Labb{\'e}}, {Franx},
  {Illingworth}, {Marchesini}, \& {Quadri}}]{Kriek09}
{Kriek}, M., {van Dokkum}, P.~G., {Labb{\'e}}, I., {Franx}, M., {Illingworth},
  G.~D., {Marchesini}, D., \& {Quadri}, R.~F. 2009, \apj, 700, 221

\bibitem[{{Lagos} {et~al.}(2016){Lagos}, {Theuns}, {Stevens}, {Cortese},
  {Padilla}, {Davis}, {Contreras}, \& {Croton}}]{Lagos16}
{Lagos}, C.~d.~P., {Theuns}, T., {Stevens}, A.~R.~H., {Cortese}, L., {Padilla},
  N.~D., {Davis}, T.~A., {Contreras}, S., \& {Croton}, D. 2016,
  ArXiv:1609.01739

\bibitem[{{Law} {et~al.}(2009){Law}, {Steidel}, {Erb}, {Larkin}, {Pettini},
  {Shapley}, \& {Wright}}]{Law09}
{Law}, D.~R., {Steidel}, C.~C., {Erb}, D.~K., {Larkin}, J.~E., {Pettini}, M.,
  {Shapley}, A.~E., \& {Wright}, S.~A. 2009, \apj, 697, 2057

\bibitem[{{Leitherer} {et~al.}(1999){Leitherer}, {Schaerer}, {Goldader},
  {Delgado}, {Robert}, {Kune}, {de Mello}, {Devost}, \&
  {Heckman}}]{Leitherer99}
{Leitherer}, C., {Schaerer}, D., {Goldader}, J.~D., {Delgado}, R.~M.~G.,
  {Robert}, C., {Kune}, D.~F., {de Mello}, D.~F., {Devost}, D., \& {Heckman},
  T.~M. 1999, \apjs, 123, 3

\bibitem[{{Ly} {et~al.}(2007){Ly}, {Malkan}, {Kashikawa}, {Shimasaku}, {Doi},
  {Nagao}, {Iye}, {Kodama}, {Morokuma}, \& {Motohara}}]{Ly07}
{Ly}, C., {Malkan}, M.~A., {Kashikawa}, N., {Shimasaku}, K., {Doi}, M.,
  {Nagao}, T., {Iye}, M., {Kodama}, T., {Morokuma}, T., \& {Motohara}, K. 2007,
  \apj, 657, 738

\bibitem[{{Mo} {et~al.}(1998){Mo}, {Mao}, \& {White}}]{Mo98}
{Mo}, H.~J., {Mao}, S., \& {White}, S.~D.~M. 1998, \mnras, 295, 319

\bibitem[{{Morishita} {et~al.}(2014){Morishita}, {Ichikawa}, \&
  {Kajisawa}}]{Morishita14}
{Morishita}, T., {Ichikawa}, T., \& {Kajisawa}, M. 2014, \apj, 785, 18

\bibitem[{{Mortlock} {et~al.}(2013){Mortlock}, {Conselice}, {Hartley},
  {Ownsworth}, {Lani}, {Bluck}, {Almaini}, \& {Duncan, K. et al.}}]{Mortlock13}
{Mortlock}, A., {Conselice}, C.~J., {Hartley}, W.~G., {Ownsworth}, J.~R.,
  {Lani}, C., {Bluck}, A.~F.~L., {Almaini}, O., \& {Duncan, K. et al.} 2013,
  \mnras, 433, 1185

\bibitem[{{Muzzin} {et~al.}(2013){Muzzin}, {Marchesini}, {Stefanon}, {Franx},
  {Milvang-Jensen}, {Dunlop}, {Fynbo}, {Brammer}, {Labb{\'e}}, \& {van
  Dokkum}}]{Muzzin13}
{Muzzin}, A., {Marchesini}, D., {Stefanon}, M., {Franx}, M., {Milvang-Jensen},
  B., {Dunlop}, J.~S., {Fynbo}, J.~P.~U., {Brammer}, G., {Labb{\'e}}, I., \&
  {van Dokkum}, P. 2013, \apjs, 206, 8

\bibitem[{{Navarro} {et~al.}(1997){Navarro}, {Frenk}, \& {White}}]{NFW}
{Navarro}, J.~F., {Frenk}, C.~S., \& {White}, S.~D.~M. 1997, \apj, 490, 493

\bibitem[{{Nelson} {et~al.}(2015){Nelson}, {van Dokkum}, {F{\"o}rster
  Schreiber}, {Franx}, {Brammer}, {Momcheva}, {Wuyts}, \&
  {Whitaker}}]{Nelson15}
{Nelson}, E.~J., {van Dokkum}, P.~G., {F{\"o}rster Schreiber}, N.~M., {Franx},
  M., {Brammer}, G.~B., {Momcheva}, I.~G., {Wuyts}, S., \& {Whitaker}, K.~E.
  e.~a. 2015, ArXiv e-prints

\bibitem[{{Newman} {et~al.}(2012){Newman}, {Genzel}, {F{\"o}rster-Schreiber},
  {Shapiro Griffin}, {Mancini}, {Lilly}, {Renzini}, \& {Bouch{\'e}, N. et
  al.}}]{Newman12}
{Newman}, S.~F., {Genzel}, R., {F{\"o}rster-Schreiber}, N.~M., {Shapiro
  Griffin}, K., {Mancini}, C., {Lilly}, S.~J., {Renzini}, A., \& {Bouch{\'e},
  N. et al.} 2012, \apj, 761, 43

\bibitem[{{Obreschkow} \& {Glazebrook}(2014)}]{Obreschkow14}
{Obreschkow}, D. \& {Glazebrook}, K. 2014, \apj, 784, 26

\bibitem[{{Obreschkow} {et~al.}(2015){Obreschkow}, {Glazebrook}, {Bassett},
  {Fisher}, {Abraham}, {Wisnioski}, \& {Green, A.~W. et al.}}]{Obreschkow15}
{Obreschkow}, D., {Glazebrook}, K., {Bassett}, R., {Fisher}, D.~B., {Abraham},
  R.~G., {Wisnioski}, E., \& {Green, A.~W. et al.} 2015, \apj, 815, 97

\bibitem[{{Oesch} {et~al.}(2010){Oesch}, {Carollo}, {Feldmann}, {Hahn},
  {Lilly}, {Sargent}, {Scarlata}, \& {Aller, et al.}}]{Oesch10}
{Oesch}, P.~A., {Carollo}, C.~M., {Feldmann}, R., {Hahn}, O., {Lilly}, S.~J.,
  {Sargent}, M.~T., {Scarlata}, C., \& {Aller, et al.} 2010, \apjl, 714, L47

\bibitem[{{Peebles}(1969)}]{Peebles69}
{Peebles}, P.~J.~E. 1969, \apj, 155, 393

\bibitem[{{Peng} {et~al.}(2002){Peng}, {Ho}, {Impey}, \& {Rix}}]{Peng02}
{Peng}, C.~Y., {Ho}, L.~C., {Impey}, C.~D., \& {Rix}, H.-W. 2002, \aj, 124, 266

\bibitem[{{Persic} \& {Salucci}(1988)}]{Persic88}
{Persic}, M. \& {Salucci}, P. 1988, \mnras, 234, 131

\bibitem[{{Puech} {et~al.}(2008){Puech}, {Flores}, {Hammer}, {Yang}, {Neichel},
  {Lehnert}, {Chemin}, \& {Nesvadba, N. et al.}}]{Puech08}
{Puech}, M., {Flores}, H., {Hammer}, F., {Yang}, Y., {Neichel}, B., {Lehnert},
  M., {Chemin}, L., \& {Nesvadba, N. et al.} 2008, \aap, 484, 173

\bibitem[{{Richard} {et~al.}(2015){Richard}, {Patricio}, {Martinez}, {Bacon},
  {Cl{\'e}ment}, {Weilbacher}, {Soto}, \& {Wisotzki, L. et al.}}]{Richard15}
{Richard}, J., {Patricio}, V., {Martinez}, J., {Bacon}, R., {Cl{\'e}ment}, B.,
  {Weilbacher}, P., {Soto}, K., \& {Wisotzki, L. et al.} 2015, \mnras, 446, L16

\bibitem[{{Roberts}(1963)}]{Roberts63}
{Roberts}, M.~S. 1963, \araa, 1, 149

\bibitem[{{Romanowsky} \& {Fall}(2012)}]{Romanowsky12}
{Romanowsky}, A.~J. \& {Fall}, S.~M. 2012, \apjs, 203, 17

\bibitem[{{Salucci} \& {Burkert}(2000)}]{Salucci00}
{Salucci}, P. \& {Burkert}, A. 2000, \apjl, 537, L9

\bibitem[{{Sandage}(1986)}]{Sandage86}
{Sandage}, A. 1986, \aap, 161, 89

\bibitem[{{Sandage} {et~al.}(1970){Sandage}, {Freeman}, \&
  {Stokes}}]{Sandage70}
{Sandage}, A., {Freeman}, K.~C., \& {Stokes}, N.~R. 1970, \apj, 160, 831

\bibitem[{{Schaller} {et~al.}(2015){Schaller}, {Frenk}, {Bower}, {Theuns},
  {Trayford}, {Crain}, {Furlong}, {Schaye}, {Dalla Vecchia}, \&
  {McCarthy}}]{Schaller15}
{Schaller}, M., {Frenk}, C.~S., {Bower}, R.~G., {Theuns}, T., {Trayford}, J.,
  {Crain}, R.~A., {Furlong}, M., {Schaye}, J., {Dalla Vecchia}, C., \&
  {McCarthy}, I.~G. 2015, \mnras, 452, 343

\bibitem[{{Schaye} {et~al.}(2015){Schaye}, {Crain}, {Bower}, {Furlong},
  {Schaller}, {Theuns}, {Dalla Vecchia}, \& {Frenk, C.~S. et al.}}]{Schaye15}
{Schaye}, J., {Crain}, R.~A., {Bower}, R.~G., {Furlong}, M., {Schaller}, M.,
  {Theuns}, T., {Dalla Vecchia}, C., \& {Frenk, C.~S. et al.} 2015, \mnras,
  446, 521

\bibitem[{{Sharples} {et~al.}(2004){Sharples}, {Bender}, {Lehnert}, {Ramsay
  Howat}, {Bremer}, {Davies}, {Genzel}, \& {Hofmann, R.}}]{Sharples04}
{Sharples}, R.~M., {Bender}, R., {Lehnert}, M.~D., {Ramsay Howat}, S.~K.,
  {Bremer}, M.~N., {Davies}, R.~L., {Genzel}, R., \& {Hofmann, R.} 2004, in
  Ground-based Instrumentation for Astronomy. Edited by Alan F. M. Moorwood and
  Iye Masanori. Proceedings of the SPIE, Volume 5492, pp. 1179-1186 (2004).,
  ed. A.~F.~M. {Moorwood} \& M.~{Iye}, 1179--1186

\bibitem[{{Sobral} {et~al.}(2009){Sobral}, {Best}, {Geach}, {Smail}, {Kurk},
  {Cirasuolo}, {Casali}, \& {Ivison, R.~J. et al.}}]{Sobral09}
{Sobral}, D., {Best}, P.~N., {Geach}, J.~E., {Smail}, I., {Kurk}, J.,
  {Cirasuolo}, M., {Casali}, M., \& {Ivison, R.~J. et al.} 2009, \mnras, 398,
  75

\bibitem[{{Sobral} {et~al.}(2015){Sobral}, {Matthee}, {Best}, {Smail},
  {Khostovan}, {Milvang-Jensen}, {Kim}, {Stott}, {Calhau}, {Nayyeri}, \&
  {Mobasher}}]{Sobral15}
{Sobral}, D., {Matthee}, J., {Best}, P.~N., {Smail}, I., {Khostovan}, A.~A.,
  {Milvang-Jensen}, B., {Kim}, J.-W., {Stott}, J., {Calhau}, J., {Nayyeri}, H.,
  \& {Mobasher}, B. 2015, \mnras, 451, 2303

\bibitem[{{Sobral} {et~al.}(2013{\natexlab{a}}){Sobral}, {Smail}, {Best},
  {Geach}, {Matsuda}, {Stott}, {Cirasuolo}, \& {Kurk}}]{Sobral13}
{Sobral}, D., {Smail}, I., {Best}, P.~N., {Geach}, J.~E., {Matsuda}, Y.,
  {Stott}, J.~P., {Cirasuolo}, M., \& {Kurk}, J. 2013{\natexlab{a}}, \mnras,
  428, 1128

\bibitem[{{Sobral} {et~al.}(2013{\natexlab{b}}){Sobral}, {Swinbank}, {Stott},
  {Matthee}, {Bower}, {Smail}, {Best}, {Geach}, \& {Sharples}}]{Sobral13b}
{Sobral}, D., {Swinbank}, A.~M., {Stott}, J.~P., {Matthee}, J., {Bower}, R.~G.,
  {Smail}, I., {Best}, P., {Geach}, J.~E., \& {Sharples}, R.~M.
  2013{\natexlab{b}}, \apj, 779, 139

\bibitem[{{Steinmetz} \& {Bartelmann}(1995)}]{Steinmetz95}
{Steinmetz}, M. \& {Bartelmann}, M. 1995, \mnras, 272, 570

\bibitem[{{Stott} {et~al.}(2014){Stott}, {Sobral}, {Swinbank}, {Smail},
  {Bower}, {Best}, {Sharples}, {Geach}, \& {Matthee}}]{Stott14}
{Stott}, J.~P., {Sobral}, D., {Swinbank}, A.~M., {Smail}, I., {Bower}, R.,
  {Best}, P.~N., {Sharples}, R.~M., {Geach}, J.~E., \& {Matthee}, J. 2014,
  ArXiv e-prints

\bibitem[{{Stott} {et~al.}(2016){Stott}, {Swinbank}, {Johnson}, {Tiley},
  {Magdis}, {Bower}, {Bunker}, \& {Harrison, C.~M. et al.}}]{Stott16}
{Stott}, J.~P., {Swinbank}, A.~M., {Johnson}, H.~L., {Tiley}, A., {Magdis}, G.,
  {Bower}, R., {Bunker}, A.~J., \& {Harrison, C.~M. et al.} 2016, \mnras, 457,
  1888

\bibitem[{{Toomre}(1964)}]{Toomre64}
{Toomre}, A. 1964, \apj, 139, 1217

\bibitem[{{Trayford} {et~al.}(2015){Trayford}, {Theuns}, {Bower}, {Schaye},
  {Furlong}, {Schaller}, {Frenk}, {Crain}, {Dalla Vecchia}, \&
  {McCarthy}}]{Trayford16}
{Trayford}, J.~W., {Theuns}, T., {Bower}, R.~G., {Schaye}, J., {Furlong}, M.,
  {Schaller}, M., {Frenk}, C.~S., {Crain}, R.~A., {Dalla Vecchia}, C., \&
  {McCarthy}, I.~G. 2015, \mnras, 452, 2879

\bibitem[{{van der Wel} {et~al.}(2014){van der Wel}, {Franx}, {van Dokkum},
  {Skelton}, {Momcheva}, {Whitaker}, {Brammer}, \& {Bell, E.~F. et
  al.}}]{vanderwel14}
{van der Wel}, A., {Franx}, M., {van Dokkum}, P.~G., {Skelton}, R.~E.,
  {Momcheva}, I.~G., {Whitaker}, K.~E., {Brammer}, G.~B., \& {Bell, E.~F. et
  al.} 2014, \apj, 788, 28

\bibitem[{{Walter} {et~al.}(2008){Walter}, {Brinks}, {de Blok}, {Bigiel},
  {Kennicutt}, {Thornley}, \& {Leroy}}]{Walter08}
{Walter}, F., {Brinks}, E., {de Blok}, W.~J.~G., {Bigiel}, F., {Kennicutt},
  Jr., R.~C., {Thornley}, M.~D., \& {Leroy}, A. 2008, \aj, 136, 2563

\bibitem[{{White}(1984)}]{White84}
{White}, S.~D.~M. 1984, \apj, 286, 38

\bibitem[{{Wisnioski} {et~al.}(2015){Wisnioski}, {F{\"o}rster Schreiber},
  {Wuyts}, {Wuyts}, {Bandara}, {Wilman}, {Genzel}, \& {Bender}}]{Wisnioski15}
{Wisnioski}, E., {F{\"o}rster Schreiber}, N.~M., {Wuyts}, S., {Wuyts}, E.,
  {Bandara}, K., {Wilman}, D., {Genzel}, R., \& {Bender}, R. e.~a. 2015, \apj,
  799, 209

\bibitem[{{Wuyts} {et~al.}(2013){Wuyts}, {F{\"o}rster Schreiber}, {Nelson},
  {van Dokkum}, {Brammer}, {Chang}, {Faber}, \& {Ferguson, H.~C. et
  al.}}]{Wuyts13}
{Wuyts}, S., {F{\"o}rster Schreiber}, N.~M., {Nelson}, E.~J., {van Dokkum},
  P.~G., {Brammer}, G., {Chang}, Y.-Y., {Faber}, S.~M., \& {Ferguson, H.~C. et
  al.} 2013, \apj, 779, 135

\bibitem[{{Zavala} {et~al.}(2016){Zavala}, {Frenk}, {Bower}, {Schaye},
  {Theuns}, {Crain}, {Trayford}, {Schaller}, \& {Furlong}}]{Zavala16}
{Zavala}, J., {Frenk}, C.~S., {Bower}, R., {Schaye}, J., {Theuns}, T., {Crain},
  R.~A., {Trayford}, J.~W., {Schaller}, M., \& {Furlong}, M. 2016, \mnras

\end{thebibliography}

\begin{landscape}
\begin{table}
\begin{minipage}{240mm} 
\caption{Galaxy Properties}
\setlength{\tabcolsep}{0.7 mm} 
\begin{tabular}{llllllllllllllllllll}
\hline                       
ID      &   RA          &  Dec               & $z$    & $V_{\rm AB}$ & $K_{\rm AB}$ & $f_{\rm neb}$ & $r_{\rm h,\star}$  &  $r_{\rm h,neb}$  & $M_{\rm H}$  & log($\frac{M_\star}{M_\odot}$) &  $\sigma_{\rm gal}$ & $\sigma_{\rm int}$ & $V$(3\,R$_{\rm d}$) & $i$  & $A_{\rm V}$   &  $j_{\star}$     &  SFR                       & Class      \\ 
     &   (J2000)  &        &      &                &                &                 & [$"$]              &  [$"$]             &              &                               &  [km\,s$^{-1}$]     & [km\,s$^{-1}$]     &  [km\,s$^{-1}$]   &      &               & [km\,s$^{-1}$kpc]  &  [M$_{\odot}$yr$^{-1}$]    &            \\
\hline
COSMOS-K1-1      &  09:59:40.603 & +02:21:04.15 & 1.6350  &  24.98  &  22.53  &  5.7   &  0.41\,$\pm$\,0.10 & 0.45\,$\pm$\,0.15 & -22.25 & 10.48 & 131\,$\pm$\,3  &  46\,$\pm$\,3  & 16\,$\pm$\,7   & 22\,$\pm$\,22 & 0.2 & 66\,$\pm$\,41 & 11 & D\\
COSMOS-K1-2      &  09:59:31.589 & +02:19:05.47 & 1.6164  &  23.70  &  20.63  &  5.1   &  0.18\,$\pm$\,0.01 & 0.24\,$\pm$\,0.05 & -24.46 & 11.40 & 176\,$\pm$\,3  &   ...          & ...            &  ...  & 0.6 & 345\,$\pm$\,6 & 16 & C\\
COSMOS-K1-3      &  09:59:33.994 & +02:20:54.58 & 1.5240  &  23.90  &  21.73  &  5.0   &  0.57\,$\pm$\,0.13 & 0.68\,$\pm$\,0.13 & -22.85 & 10.57 & 123\,$\pm$\,3  &  38\,$\pm$\,3  & 206\,$\pm$\,30 & 37\,$\pm$\,10 & 0.4 & 1171\,$\pm$\,463 & 10 & D\\
COSMOS-K1-4      &  09:59:28.339 & +02:19:50.53 & 1.4855  &  21.02  &  19.96  &  86.9  &  0.08\,$\pm$\,0.05 & 0.30\,$\pm$\,0.26 & -24.51 & 11.09 & 580\,$\pm$\,3  &   ...          & ...            &  ...  & 0.0 &  ...  & 100 & I\\
COSMOS-K1-10     &  09:59:30.902 & +02:18:53.04 & 1.5489  &  24.06  &  21.60  &  15.4  &  0.20\,$\pm$\,0.02 & 0.04\,$\pm$\,0.05 & -23.21 & 10.72 & 159\,$\pm$\,3  &   ...          & ...            &  ...  & 0.6 & 344\,$\pm$\,5 & 44 & C\\

\hline
\end{tabular}
\label{tab:props}
\end{minipage} 
\noindent{\footnotesize The full table is given in the online version
  of this paper.  The first five rows are shown here for their
  content.  Notes: $f_{\rm neb}$ denotes the nebular emission line
  flux ([O{\sc ii}] in the case of MUSE and H$\alpha$ for KMOS) in
  units of 10$^{-17}$\,erg\,cm$^{-2}$\,s$^{-1}$.  $r_{\rm h,\star}$
  and $r_{\rm h,neb}$ are the (deconvolved) continuum and nebular
  emission half light radii respectively.  $\sigma_{\rm neb}$ denotes
  the galaxy-integrated velocity dispersion as measured from the
  one-dimensional spectrum.  $\sigma_{\rm int}$ denotes the average
  intrinsic velocity dispersion within the galaxy (after correcting
  for beam smearing effects).  $V$(3\,R$_{\rm d}$) is the observed
  velocity at 3\,R$_{\rm d}$.  $i$ is the disk inclination.  SFR is
  measured from the [O{\sc ii}] flux with
  SFR\,=\,0.8\,$\times$\,10$^{-41}$\,L$_{\rm [OII]}$\,erg\,s$^{-1}$
  and correcting for dust reddenning using the Calzetti redenning law.
  \\}
\end{table}
\end{landscape}


\end{document}